\documentclass[reprint,aps,prd,superscriptaddress,nofootinbib,floats,showpacs]{revtex4-2}

\usepackage{hyperref}
\usepackage{graphicx}
\usepackage{mathtools}
\usepackage{amsmath,amssymb,amsfonts,amsthm,latexsym,stmaryrd,physics,mathrsfs}
\allowdisplaybreaks[4]
\usepackage{marginnote}
\usepackage{color}
\usepackage{bm}
\usepackage{float}
\usepackage{nicefrac}
\usepackage{microtype} 
\usepackage{verbatim}
\usepackage{setspace}
\usepackage[T1]{fontenc}
\usepackage[utf8]{inputenc}
\usepackage{dsfont}
\usepackage{tikz}
\usepackage{geometry}
\usepackage{enumitem}

\begin{document}

\title{Loop Quantum Vector-Tensor Gravity and Its Spherically Symmetric Model}

\author{Shengzhi Li}
\affiliation{School of Physics and Technology, Xinjiang University, Urumqi 830046, China}
\affiliation{School of Physics and Astronomy, Key Laboratory of Multiscale Spin Physics (Ministry of Education), Beijing Normal University, Beijing 100875, China}

\author{Yongge Ma}
\email{Contact author: mayg@bnu.edu.cn}
\affiliation{School of Physics and Technology, Xinjiang University, Urumqi 830046, China}
\affiliation{School of Physics and Astronomy, Key Laboratory of Multiscale Spin Physics (Ministry of Education), Beijing Normal University, Beijing 100875, China}

\begin{abstract}
	The Hamiltoinian analysis of the vector-tensor theory of gravity is performed. The resulting geometrical dynamics is reformulated into the connection dynamics, with the real $\mathrm{SU}(2)$-connection serving as one of the configuration variables. This formulation allows us to extend the loop quantization scheme of general relativity to the vector-tensor theory, thereby rigorously constructing its quantum kinematical framework. The scalar constraint is promoted to a well-defined operator in the vertex Hilbert space, to represent quantum dynamics. Moreover, the spherically symmetric model of the vector-tensor theory is obtained by the symmetric reduction. Following the general deparametrization strategy for theories with diffeomorphism invariance, the spherically symmetric model can be fully deparametrized in terms of the degrees of freedom of the vector field. The corresponding reduced phase space quantization is carried out. The physical Hamiltonian generating relative evolution is promoted to a well-defined operator on the physical Hilbert space.
\end{abstract}

\maketitle

\section{Introduction}\label{section1}
As a background-independent and non-perturbative quantization of general relativity (GR), loop quantum gravity (LQG) has been widely studied in recent decades \cite{Ashtekar1, Thiemann,Rovelli:2004tv, Muxin}. The key to the successful execution of this quantization approach lies in the fact that classical GR can be recast in the form of $\mathrm{SU}(2)$-connection dynamics. Thus, it is desirable to explore whether other theories of gravity possess this feature as well. It has been shown in previous study that the metric $f(R)$ theory and scalar-tensor theory can also be formulated into connection dynamical formalism and hence be quantized by the method of LQG \cite{Xiangdong1, Zhang:2011qq, Xiangdong2, Zhang:2011gn}. Recently the vector-tensor theory of gravity has garnered much attention due to its potential to address cosmological issues such as dark energy \cite{BeltranJimenez:2008iye, BeltranJimenez:2010nxz, Koivisto:2008xf} and dark matter \cite{Zlosnik:2006zu, Zhao:2007ce}. In addition to the metric field, the theory also introduces a dynamical vector field to describe gravity \cite{Will:1972zz, Hellings:1973zz}. In this paper, we are going to perform the Hamiltonian analysis of the vector-tensor gravity and derive the connection dynamics of the vector-tensor theory from its geometric dynamics by performing a canonical transformation and extending the phase space. The connection-dynamical system is subject to the Gauss, vector, and scalar constraints, together forming a first-class constraint system. By applying the methods of LQG, the Gauss and vector constraints can be solved at the quantum level, and the scalar constraints can be promoted to a well-defined operator on the vertex Hilbert space \cite{Jinsong, Lewandowski1,Alesci:2015wla}. 

In gravity theories with diffeomorphism invariance similar to GR, the Hamiltonian is a linear combination of first-class constraints, generating gauge transformations. This implies that the Dirac observables, i.e., gauge invariants, do not evolve. This theoretical result clearly contradicts the real world. Therefore, it is essential to understand the meaning of evolution of Dirac observables. To address this issue, the concept of deparametrization was proposed for GR \cite{Brown, Rovelli1, Rovelli2, Dapor:2013hca, Domagala:2010bm,Dittrich1, Dittrich2, Thiemann1, Giesel1, Giesel2, Giesel6, Thiemann3, Giesel4}, by introducing dynamical matter fields as a physical reference system. The corresponding Dirac observables can be constructed, along with the physical Hamiltonian that generates the evolution of these observables with respect to the physical time. While the deparametrization program can be successfully applied to GR coupled with suitable matters field, it remains open how to deparametrize a pure gravity theory. An idea in the present work is to study whether the vector-tensor theory can deparametrized using the vector field of gravity itself, thereby addressing the evolution issue of the Dirac observables in this theory in the vacuum case. For simplicity, we will study this issue in the spherically symmetric model. It turns out that the degrees of freedom of the vector field can indeed be used as the radial and time coordinates to deparametrize the theory. The reduced phase space with all constraints resolved and a time-dependent physical Hamiltonian can be obtained. Moreover, by performing loop quantization on the reduced phase space, we circumvent the difficulty of solving the scalar constraint at the quantum level, obtain the physical Hilbert space, and promote the physical Hamiltonian to a well-defined operator on it.

The structure of this paper is as follows. In Section \ref{LQVTGT}, we first perform a Hamiltonian analysis of the vector-tensor theory, and then derive its connection dynamics. Based on the connection formalism, we carry out the loop quantization of the vector-tensor theory. In Section \ref{DVTGSSM}, we first review the general process of deparametrizing the theories with diffeomorphism invariance using spacetime scalar fields, then compare the dynamical results of gauge-fixed theories and deparametrized theories. Finally, the spherically symmetric model of the vector-tensor theory is deparametrized by using the vector field degrees of freedom. In Section \ref{QRPSVTGT}, we perform loop quantization on the reduced phase space to obtain the physical Hilbert space and promote the physical Hamiltonian to an operator on it. Finally, in Section \ref{CR}, we summarize our results and draw conclusions.

\section{Connection Dynamics and Loop Quantization}\label{LQVTGT}
\subsection{Hamiltonian Analysis}
The vector-tensor theory of gravity without the potential term is given by the action \cite{BeltranJimenez:2008zzi,Will2}:
\begin{align}\label{action}
S[g,Y]\!&=\!\int_{\mathcal{M}}\!d^4x\sqrt{-g}[-\frac{1}{16\pi G}R\!+\!\varpi RY_{\mu}Y^{\mu}\nonumber\\
&+\rho\Omega_{\mu\nu}\Omega^{\mu\nu}\!+\!\varkappa{R_{\mu\nu}Y^{\mu}Y^{\nu}}\!+\!\varsigma\nabla_{\mu}Y_{\nu}\nabla^{\mu}Y^{\nu}],
\end{align}
where $\varpi, \rho, \varkappa, \varsigma$ are dimensionless coupling parameters, $g_{\mu\nu}$ is the spacetime metric, $\nabla_{\mu}$ is the derivative operator compatible with $g_{\mu\nu}$, $R_{\mu\nu}$ is the Ricci tensor, $R\equiv\!g^{\mu\nu}R_{\mu\nu}$, $g\equiv\det(g_{\mu\nu})$, $Y_{\mu}$ is a timelike vector field, and $\Omega_{\mu\nu}\equiv\partial_{\mu}Y_{\nu}-\partial_{\nu}Y_{\mu}$. The spacetime indices $\mu,\nu,\cdots= 0,1,2,3$ are raised and lowered by $g_{\mu\nu}$. In this paper, we consider the case where $\varkappa=\varsigma=0$. For convenience, we absorb the minus sign of the first term into the dimensionless parameters and set $\omega=16\pi G\varpi$.

In the Hamiltonian formulation, one assumes that the spacetime can be decomposed as $\mathcal{M}=\mathbb{R}\times\Sigma$, where $\mathbb{R}$ is the time dimension with coordinate $t$, and $\Sigma$ is a 3-dimensional spatial manifold with coordinates $x^{a}$, where $a,b,\cdots=1,2,3$. Let $n^{\mu}$ be the future-directed normal vector to the spatial hypersurface $\Sigma$ with respect to $t$. Then, the spatial metric induced by $g_{\mu\nu}$ on  $\Sigma$ is $q_{\mu\nu}=g_{\mu\nu}+n_{\mu}n_{\nu}$, which corresponds to the spatial tensor $q_{ab}$. Thus, the spacetime metric $g_{\mu\nu}$ can be decomposed as:
\begin{equation}
	g_{\mu\nu}dx^{\mu}dx^{\nu}\!=\!-N^2dt^2\!+\!q_{ab}(dx^{a}\!+\!N^{a}dt)(dx^{b}\!+\!N^{b}dt),
\end{equation}
where $N$ is the lapse function and $N^{a}$ is the shift vector. The scalar curvature $R$ can be expressed as:
\begin{equation}
	R\!={}^{(3)}\!R\!+\!K_{ab}K^{ab}\!-\!K^2\!-\!2\nabla_{\mu}(n^{\nu}\nabla_{\nu}n^{\mu}\!-\!n^{\mu}K),
\end{equation}
where, $K_{ab}$ is the extrinsic curvature of $\Sigma$, $^{(3)}\!R$ is the scalar curvature of the spatial metric $q_{ab}$, and $K\equiv\!K_{ab}q^{ab}$. The spatial indices are raised and lowered by $q_{ab}$.

The future-directed timelike vector field $Y_{\mu}$ can be decomposed as:
\begin{equation}
	Y_{\mu}dx^{\mu}=\sqrt{q^{ab}Y_{a}Y_{b}-\check{Y}}Ndt+Y_{a}(dx^{a}+N^{a}dt),
\end{equation}
where, $Y_{a}$ is the spatial projection of $Y_{\mu}$, and $\check{Y}\equiv\!Y_{\mu}Y^{\mu}$ is a spacetime scalar field. The third term in the action (\ref{action}) can be expressed as:
\begin{equation}
	\Omega_{\mu\nu}\Omega^{\mu\nu}=\Omega_{ab}\Omega^{ab}-2\mathcal{E}_{a}\mathcal{E}^{a},
\end{equation}
where, $\mathcal{E}_{a}$ is the spatial tensor corresponding to $\mathcal{E}_{\mu}\equiv(\delta_{\mu}^{\nu}+n_{\mu}n^{\nu})n^{\sigma}\Omega_{\nu\sigma}$, and $\Omega_{ab}$ is the spatial projection of $\Omega_{\mu\nu}$.

Collecting the above results, by performing integration by parts and neglecting boundary terms, the action (\ref{action}) can be written as:
\begin{align}
		S[g,Y]\!&=\!\int_{\mathbb{R}}\!dt\!\int_{\Sigma}\!dx^{3}\!N\sqrt{q}[\frac{1\!+\!\omega \check{Y}}{16\pi G}({}^{(3)}\!R\!+\!K_{ab}K^{ab}\nonumber\\
		&-K^2)+\frac{\omega}{8\pi G}(\frac{D_{a}N}{N}D^{a}\check{Y}-K\partial_{n}\check{Y})\nonumber\\
		&+\rho(\Omega_{ab}\Omega^{ab}-2\mathcal{E}_{a}\mathcal{E}^{a})]\nonumber\\
		&=:\int_{\mathbb{R}}dt\int_{\Sigma}dx^{3}\mathcal{L},
\end{align}
where $D_{a}$ is the derivative operator compatible with $q_{ab}$, $q\equiv\det(q_{ab})$, and $\partial_{n}\check{Y}\equiv\!n^{\mu}\nabla_{\mu}\check{Y}$. We can choose $N, N^{a}, \check{Y}, Y_{a}, q_{ab}$ as the configuration variables, and their corresponding conjugate momenta respectively read:
\begin{equation}\label{momentum}
	\begin{split}
		\pi_N&\!:=\!\frac{\delta{S}}{\delta(\partial_{t}N)}\!=\!0,\quad\,
		\pi^{Y}\!:=\!\frac{\delta{S}}{\delta(\partial_{t}\check{Y})}\!=\!-\frac{\omega q^{1/2}K}{8\pi G},\\
		\pi_{N^a}&\!:=\!\frac{\delta{S}}{\delta(\partial_{t}{N}^a)}\!=\!0,\quad
		\pi^a\!:=\!\frac{\delta{S}}{\delta(\partial_{t}{Y}_{a})}\!=\!4\rho q^{1/2}\mathcal{E}^{a},\\
		p^{ab}&\!:=\!\frac{\delta{S}}{\delta(\partial_{t}{q}_{ab})}\!=\!\frac{q^{1/2}}{16\pi G}((1\!+\!\omega \check{Y})(K^{ab}\!-\!Kq^{ab})\\
		&-\omega q^{ab}\partial_n\check{Y}),
	\end{split}
\end{equation}
where $\pi_N=\pi_{N^a}=0$ give the primary constraints, and $\partial_{t}{\check{Y}}, \partial_{t}{Y}_{a}, \partial_{t}{q}_{ab}$ can be solved by the remaining equations. The non-trivial Poisson brackets of the fundamental variables read:
\begin{equation}
	\begin{split}
		\{N(x),\pi_{N}(y)\}&=\delta^{(3)}(x,y),\\
		\{\check{Y}(x),\pi^{Y}(y)\}&=\delta^{(3)}(x,y),\\
		\{N^{a}(x),\pi_{N^{b}}(y)\}&=\delta^{a}_{b}\delta^{(3)}(x,y),\\
		\{Y_{a}(x),\pi^{b}(y)\}&=\delta^{b}_{a}\delta^{(3)}(x,y),\\
		\{q_{ab}(x),p^{cd}(y)\}&=\delta^{(c}_a\delta^{d)}_b\delta^{(3)}(x,y).
	\end{split}
\end{equation}
The canonical Hamiltonian of the system reads:
\begin{align}
	H_{can}\!:=&\!\int_{\Sigma}dx^{3}\pi^{Y}\partial_{t}{\check{Y}}+\pi^a\partial_{t}{Y}_{a}+p^{ab}\partial_{t}{q}_{ab}-\mathcal{L}\nonumber\\
	=&\int_{\Sigma}dx^{3}NC+N^{a}C_{a},
\end{align}	
where the scalar constraint $C$ and the diffeomorphism constraint $C_{a}$ are defined by 
\begin{align}
	C&\!:=\!\frac{16\pi Gq^{-1/2}}{1\!+\!\omega \check{Y}}(p_{ab}p^{ab}\!-\!\frac{p^2}{3})\!-\!\frac{16\pi G}{3\omega}q^{-1/2}p\pi^{Y}\nonumber\\
	&-\frac{(1\!+\!\omega \check{Y})q^{1/2}}{16\pi G}{^{(3)}R}\!-\!\rho q^{1/2}\Omega_{ab}\Omega^{ab}\!-\!\frac{\pi_{a}\pi^{a}}{8\rho q^{1/2}}\nonumber\\
	&+\!\frac{8\pi{G}(\pi^{Y})^{2}}{3\omega^{2}q^{1/2}}(1\!+\!\omega \check{Y})\!+\!\frac{\omega q^{1/2}}{8\pi G}D_{a}D^{a}\check{Y}\nonumber\\
	&+(q^{ab}Y_{a}Y_{b}-\check{Y})^{1/2}D_{c}\pi^{c},\label{scalar constraint}\\
	C_{a}&\!:=\!-2q_{ab}D_{c}p^{bc}\!+\!\pi^{Y}D_{a}\check{Y}\!-\!\pi^{b}\Omega_{ba}\!-\!Y_{a}D_{b}\pi^{b}.\label{diffeomorphism constraint}
\end{align}
with $p\equiv p^{ab}q_{ab}$. They are the secondary constraints that follow from the consistency condition of the primary constraints. Their smeared version read $C[N]=\int_{\Sigma}dx^{3}NC$, and  $\vec{C}[\vec{N}]=\int_{\Sigma}dx^{3}N^{a}C_{a}$. The Poisson brackets between them can be calculated as (Appendix \ref{appendix}):
\begin{equation}\label{constraint algebra}
	\begin{split}
		&\{\vec{C}[\vec{N}],\vec{C}[\vec{M}]\}=C[\mathscr{L}_{\vec{N}}\vec{M}],\\
		&\{\vec{C}[\vec{N}],C[M]\}=C[\mathscr{L}_{\vec{N}}M],\\
		&\{C[N],C[M]\}=\vec{C}[ND^aM-MD^aN],
	\end{split}
\end{equation}
where $\mathscr{L}$ denotes the Lie derivative. Therefore, this is a first-class constraint system with no further constraints. We can naturally  resolve the primary constraints, leading to a phase space of geometric dynamics formed by the remaining fundamental variables and secondary constraints.

\subsection{Connection Dynamics}

To perform loop quantization, one needs to extend the phase space and introduce the connection-dynamical formulation \cite{Xiangdong1, Zhang:2011qq, Xiangdong2, Zhang:2011gn,Li:2022dei}. It is convenient to define 
\begin{align}
	\tilde{K}^{ab}:=&16\pi Gq^{-1/2}(p^{ab}-\frac{1}{2}pq^{ab})\nonumber\\
	=&(1+\omega \check{Y})K^{ab}+\frac{\omega}{2}q^{ab}\partial_{n}\check{Y},
\end{align}
which can play the same role as the extrinsic curvature in connection formulation of GR. Then, the geometric variables of the connection dynamics can be defined as
\begin{equation}
	A^{i}_{a}:=\Gamma^{i}_{a}+\gamma\tilde{K}^{i}_{a},\qquad E^{a}_{i}:=q^{1/2}e^{a}_{i},
\end{equation} 
where $e^{a}_{i}$ is the triad such that $q_{ab}e^{a}_{i}e^{b}_{j}=\delta_{ij}$, $\Gamma^{i}_{a}$ is the spin connection defined by $E^{a}_{i}$, $\gamma$ is a nonzero real number, and $\tilde{K}^i_a\equiv\tilde{K}_{ab}e^{b}_{j}\delta^{ij}$. The internal indices $i,j =1,2,3$ are raised and lowered by $\delta_{ij}$. It is easy to see that the initial geometric variables $q_{ab}$ and $p^{ab}$ can be expressed in terms of $A^{i}_{a}$ and $E^{a}_{i}$. By setting the non-trivial Poisson bracket between $A^{i}_{a}$ and $E^{a}_{i}$ as 
\begin{equation}\label{AE}
	\{A^i_{a}(x),E^{b}_j(y)\}=8\pi G\gamma\delta^{b}_a\delta^{i}_j\delta^{(3)}(x,y), 
\end{equation}
the symplectic structure of the initial geometric variables can be obtained from (\ref{AE}) by the symplectic reduction \cite{Thiemann}. We thus obtain an extended phase space. To ensure that the extended phase space can be reduced back to the original phase space, we introduced the following Gauss constraint:
\begin{equation}\label{Gaussian constraint}
	G_{i}=\frac{1}{8\pi\gamma G}(\partial_aE^a_i+\epsilon_{ij}{}^{k}A^j_aE^{a}_{k}).
\end{equation} 
The original diffeomorphism and scalar constraints can be expressed in terms of the connection dynamical variables as
\begin{align}
C&=\frac{1\!+\!\omega \check{Y}}{16\pi G}q^{-\frac{1}{2}}\epsilon_{i}{}^{jk}F^{i}_{ab}E^{a}_{j}E^{b}_{k}\!+\!\frac{(2/3)q^{-\frac{1}{2}}(\tilde{K}^{i}_{a}E^{a}_{i})^{2}}{16\pi G(1\!+\!\omega \check{Y})}\nonumber\\
&-\frac{1}{8\pi G}(\dfrac{1}{1\!+\!\omega \check{Y}}\!+\!(1\!+\!\omega \check{Y})\gamma^{2})q^{-\frac{1}{2}}\tilde{K}^{[i}_{a}\tilde{K}^{j]}_{b}E^{a}_{i}E^{b}_{j}\nonumber\\
&+\!\frac{2}{3\omega}q^{\!-\!1/2}\tilde{K}^{i}_{a}E^{a}_{i}\pi^{Y}\!+\!\frac{8\pi G}{3\omega^{2}}(1\!+\!\omega \check{Y})q^{\!-\!1/2}(\pi^{Y})^{2}\nonumber\\
&-\frac{\pi_{a}\pi^{a}}{8\rho q^{\frac{1}{2}}}-\rho q^{\frac{1}{2}}\Omega_{ab}\Omega^{ab}+\!\frac{\omega q^{\frac{1}{2}}}{8\pi G}\!D_{a}D^{a}\check{Y}\nonumber\\
&+\frac{D_{c}\pi^{c}}{q^{\frac{1}{2}}}(E^{ai}Y_{a}E^{b}_{j}Y_{b}-q\check{Y})^{1/2}\label{constraint_C},\\
C_{a}&=\frac{1}{8\pi\gamma G}F^{i}_{ab}E^{b}_{i}-A^{i}_{a}G_{i}+\pi^{Y}D_{a}\check{Y}-\pi^{b}\Omega_{ba}\nonumber\\
&-Y_{a}D_{b}\pi^{b},\label{constraint_Ca}
\end{align}
where $F^{i}_{ab}\equiv2\partial_{[a}A^{i}_{b]}+\epsilon^{i}{}_{jk}A^{j}_{a}A^{k}_{b}$. The three kinds of constraints form a first-class constraint system.

\subsection{Kinematic Hilbert Space}
The phase space of connection dynamics in the vector-tensor theory consists of geometric variables $(A^{i}_{a},E^{b}_{j})$, vector field variables $(Y_{a},\pi^{b})$, and scalar field variables $(\check{Y},\pi^{Y})$. Therefore, the kinematic Hilbert space of the theory is the direct product of these three parts. Now, we perform loop quantization for each of them.

In the geometric sector, given a graph ${\alpha}=\{e_{1},\cdots,e_{n}\}$, we define $\mathrm{Cyl}_{\alpha}$ as the vector space generated by finite linear combinations of functions of the following form \cite{Ashtekar1,Rovelli:2004tv,Thiemann,Muxin}:
\begin{equation}
	\Psi_{\alpha}(A)=\psi(h_{e_{1}}(A),\cdots,h_{e_{n}}(A)),
\end{equation}
where $\psi$ is a smooth function from $\mathrm{SU}(2)^n$ to $\mathbb{C}$, and $h_{e}(A)\equiv\mathcal{P}\exp(-\int_{e}A_{a}^{i}\tau_{i})$ is the holonomy of $A^i_a$ along edge $e$. Here the $SU(2)$ generators are given by $\tau_{j}={\sigma_{j}}/{2i}$, with $\sigma_{j}$ being the Pauli matrix. Since we can always find a graph $\alpha''=\{e''_{1},\cdots,e''_{n''}\}$ such that $\Psi_{\alpha}$ and $\Psi'_{\alpha'}$ can be viewed as elements of $\mathrm{Cyl}_{\alpha''}$, we define the inner product on $\mathrm{Cyl}_{geo}\equiv\cup_{\alpha}\mathrm{Cyl}_{\alpha}$ as:
\begin{align}
	\left<\Psi_{\alpha},\Psi'_{\alpha'}\right>_{kin}:=&\int_{SU(2)^{n''}}{dh_{1},\cdots,dh_{n''}}\nonumber\\
	\times&\overline{\psi(h_{1},\cdots,h_{n''})}\psi'(h_{1},\cdots,h_{n''}),\label{scalar product1}
\end{align}
where the integral is with respect to the Haar measure on $\mathrm{SU}(2)^{n''}$. By completing $\mathrm{Cyl}_{geo}$ with the inner product (\ref{scalar product1}), we obtain the kinematic Hilbert space $\mathcal{H}_{kin,geo}$ for the geometric sector. It has an orthonormal basis $\{T_{\tilde{s}}\}$, with $T_{\tilde{s}}(A)=1$ when $\tilde{s}=\emptyset$, and otherwise \cite{Ashtekar1,Thiemann,Rovelli:2004tv,Muxin,Ilkka}:
\begin{align}
	T_{\tilde{s}}(A)=&(\prod\limits_{v\in{V(\alpha)}}(i_{v})^{n_{e_{1}}\cdots n_{e_{N_{s}}}}_{m_{e'_{1}}\cdots m_{e'_{N_{t}}}})\nonumber\\
	\times&(\prod\limits_{e\in{E(\alpha)}}\sqrt{2j_{e}+1}(h^{(j_{e})}_{e}(A))^{m_{e}}_{n_{e}}),
\end{align}
where $\tilde{s}\equiv(\alpha,\vec{j},\vec{i})$, $\alpha$ ranges over all unoriented graphs, and thereafter, we fix their orientations once and for all, $E(\alpha)$ and $V(\alpha)$ denote the sets of edges and vertices in $\alpha$ respectively, $j_{e}\ne0$ labels the irreducible representation spaces $\mathcal{H}_{j_{e}}$ of $\mathrm{SU}(2)$, and $(h^{(j)}_{e})^{m}_{n}\equiv{\left<jm\right|}h_{e}{\left|jn\right>}$ is the Wigner-D matrix, acting as the multiplication operator on $\mathcal{H}_{kin,geo}$, $e_{1},\cdots,e_{N_{s}}$ are the edges with the beginning point $b(e)=v$, $e'_{1},\cdots,e'_{N_{t}}$ are the edges with the final point $f(e')=v$, and $\{i_{v}\}$ is the basis obtained by decomposing $(\otimes_{b(e)=v}\mathcal{H}_{j_{e}})\otimes(\otimes_{f(e')=v}\mathcal{H}^{*}_{j_{e'}})$ into a direct sum of irreducible representations, which is required to be non-trivial at the pseudo-vertex.

It is obvious that the construction of $\mathcal{H}_{kin,geo}$ parallels that of standard LQG. Therefore, we can similarly introduce the operator corresponding to the Gauss constraint (\ref{Gaussian constraint}) on it. By restricting $\alpha$ and $i_{v}$ in $T_{\tilde{s}}$ to graphs without pseudo-vertices and to intertwiners respectively, we obtain $\mathrm{SU}(2)$ gauge-invariant $T_{s}$, which forms an orthonormal basis in the kernel space $\mathcal{H}^{G}_{kin,geo}$ of the Gauss constraint operator \cite{Ashtekar1,Thiemann,Rovelli:2004tv,Muxin,Ilkka}. Furthermore, the flux operators for $E^{a}_{i}$ and the spatial geometry operators, such as the volume operator $\hat{V}$, can also be introduced\cite{Ashtekar4}.

Analogous to the treatment in Ref.\cite{Thiemann:1997rt}, in the vector field sector, given a graph ${\beta}=\{e_{1},\cdots,e_{n}\}$, we define $\mathrm{Cyl}_{\beta}$ as the vector space generated by finite linear combinations of functions of the following form:
\begin{equation}
	T_{u}(Y)=\prod\limits_{e\in{E(\beta)}}\exp(i\zeta_{e}\int_{e}Y_{a}),
\end{equation}
where $u\equiv(\beta,\vec{\zeta})$, $E(\beta)$ denotes the set of edges in $\beta$, and $\zeta_{e}$ is an arbitrary real number. Because it is always possible to find a graph $\beta''=\{e''_{1},\cdots,e''_{n''}\}$ such that $T_{u}$ and $T_{u'}$ can be interpreted as elements of $\mathrm{Cyl}_{\beta''}$, we define the inner product on $\mathrm{Cyl}_{\mathcal{Y}}\equiv\cup_{\beta}\mathrm{Cyl}_{\beta}$ as:
\begin{equation}\label{scalar product2}
	\left<T_{u},T_{u'}\right>_{kin}:=\prod\limits_{e''\in{E(\beta'')}}\delta_{\zeta_{e''},\zeta'_{e''}},
\end{equation}
where the Kronecker-$\delta$ symbol is used. By completing $\mathrm{Cyl}_{\mathcal{Y}}$ using the inner product (\ref{scalar product2}), we get the kinematic Hilbert space $\mathcal{H}_{kin,\mathcal{Y}}$ associated with the vector field sector. Let ${\beta}$ range over all unoriented graphs, and thereafter fix their orientations once and for all. Then, $\{T_{u}\}$ forms an orthonormal basis, where $T_{u}=1$ when $u=\emptyset$, otherwise $\zeta_{e}\ne0$ on each edge, and each pseudo-vertex satisfies the non-trivial condition: for two edges $e$ and $e'$ at the pseudo-vertex $v$, if $b(e)=f(e')=v$, then $\zeta_{e}\ne\zeta_{e'}$; if $b(e)=b(e')=v$ or $f(e)=f(e')=v$, then $\zeta_{e}\ne-\zeta_{e'}$.

The operator corresponding to the flux, $\pi[S]\equiv\int_{S}d^{2}\sigma n_{a}(\sigma)\pi^{a}(\sigma)$, of $\pi^{a}$ through a surface $S$ with normal vector $n_{a}$ acts on the basis as 
\begin{equation}\label{flux}
	\hat{\pi}[S]\cdot T_{u}=\frac{\hbar}{2}\sum_{v\in{V(\beta)}\cap{S}}\sum_{e|_{v}\in{E(\beta)}}\kappa(S,e)\zeta_{e}T_{u},
\end{equation}
where, $T_{u}$ is adjusted to a form compatible with the surface $S$ before being acted, such that all intersections of surface $S$ with the graph $\beta$ are vertices of $\beta$, and they serve as the beginning points of edges $e\in E(\beta)$ intersecting with $S$, $e|_{v}$ denotes an edge with endpoint $v$, and when the orientation of edge $e$ matches or opposes the orientation of surface $S$, $\kappa(S,e)=\pm 1$; otherwise, $\kappa(S,e)=0$.

In the scalar field sector, we define $\mathrm{Cyl}_{\mathcal{S}}$ as the vector space generated by finite linear combinations of functions of the following form \cite{Lewandowski:2011xf,Thiemann:1997rt,Ashtekar:2002vh}:
\begin{equation}
	T_{c}(\check{Y})=\exp(i\sum\limits_{x\in\Sigma}c(x)\check{Y}(x)),
\end{equation}
where $c:\Sigma\to\mathbb{R}$ is an arbitrary function with finite support. Then, the inner product on $\mathrm{Cyl}_{\mathcal{S}}$ can be defined as:
\begin{equation}\label{scalar product3}
	\left<T_{c},T_{c'}\right>_{kin}=\delta_{c,c'}.
\end{equation}
By completing $\mathrm{Cyl}_{\mathcal{S}}$ with (\ref{scalar product3}), we acquire the kinematic Hilbert space $\mathcal{H}_{kin,\mathcal{S}}$ for the scalar field sector, with $\{T_{c}\}$ forming an orthonormal basis.

Moreover, we can promote the momentum $\int_{\Sigma}d^3xf(x)\pi^{Y}(x)$ to a well-defined operator acting on the basis as
\begin{equation}
	\int_{\Sigma}d^3xf(x)\hat{\pi}^{Y}(x)\cdot T_{c}=\hbar\sum_{v\in{X}}f(v)c_{v}T_{c},
\end{equation}
where $X\equiv\!supp(c)$, $c_{v}\equiv\!c(v)$, and as the index of $T_{c}$, $c$ can be viewed as $c=(X,\vec{c}:=\{c_{v}\})$.

\subsection{$\mathrm{SU}(2)$ Gauge-Invariant and Diffemorphism-Invariant Hilbert Space}
Since the Gauss constraint only involves the geometric sector, we can directly obtain the $\mathrm{SU}(2)$ gauge-invariant Hilbert space as $\mathcal{H}^{G}_{kin}\equiv\mathcal{H}^{G}_{kin,geo}\otimes\mathcal{H}_{kin,\mathcal{Y}}\otimes\mathcal{H}_{kin,\mathcal{S}}$. Using the basis vectors of their respective Hilbert spaces, we obtain an orthonormal basis $\{T_{s,u,c}\equiv\!T_{s}T_{u}T_{c}\}$ for $\mathcal{H}^{G}_{kin}$.

Any semi-analytic diffeomorphism $\varphi\in Diff$ can be promoted to the following unitary operator:
\begin{equation}\label{diffeomorphism}
	\hat{U}(\varphi)\!\cdot\! T_{s,u,c}(A,Y,\check{Y})\!:=\!T_{s,u,c}(\varphi^{*}A,\varphi^{*}Y,\varphi^{*}\check{Y}),
\end{equation}
where $\varphi^{*}$ denotes the pullback map. To obtain the diffeomorphism-invariant Hilbert space, one can use the group-averaging method \cite{Ashtekar5,Giulini:1998rk,Giulini:1998kf}. We define $\mathcal{D}^{G}_{kin}\equiv\!span_{\mathbb{C}}\{T_{s,u,c}\}$. Then the anti-linear rigging map $\eta:\mathcal{D}^{G}_{kin}\to(\mathcal{D}^{G}_{kin})^{*}$ is defined as 
\begin{equation}
	\eta(T_{s,u,c})\!:=\!\sum_{\varphi\in{D\!i\!f\!f/T\!D\!i\!f\!f_{\alpha,\beta,X}}} \big<\hat{U}(\varphi)\!\cdot\!T_{s,u,c},\cdot\big>_{kin},
\end{equation}
where $TDiff_{\alpha,\beta,X}\equiv\{\varphi\in{Diff}\mid\varphi(e)=e,\varphi(v)=v,\forall{e\in{E(\alpha)}\cup{E(\beta)}},\forall{v\in{X}}\}$, and $\varphi$ is a representative map which is fixed once and for all within each coset. We can define the inner product on $\eta[\mathcal{D}^{G}_{kin}]$ as
\begin{equation}\label{scalar product4}
	\left<\eta(T_{s,u,c}),\eta(T_{s',u',c'})\right>_{Diff}:=[\eta(T_{s,u,c})](T_{s',u',c'}).
\end{equation}
By completing $\eta[\mathcal{D}^{G}_{kin}]$ with (\ref{scalar product4}), the $\mathrm{SU}(2)$ gauge-invariant and diffeomorphism-invariant Hilbert space $\mathcal{H}^G_{Diff}$ can be obtained.

\subsection{Scalar Constraint Operator}
Now, we define the scalar constraint as an operator on $\mathcal{H}^{G}_{kin}$ through appropriate regularization. For convenience, we divide the scalar constraint $C[N]$ into the following 9 terms to be treated separately:
\begin{align}
		C_{1}[N]=&\int_{\Sigma}{d^3x}N\frac{1+\omega \check{Y}}{16\pi G}q^{-\frac{1}{2}}\epsilon_{i}^{\,jk}F^{i}_{ab}E^{a}_{j}E^{b}_{k},\nonumber\\
		C_{2}[N]=&\int_{\Sigma}{d^3x}\frac{-N}{8\pi G}(\dfrac{1}{1\!+\!\omega \check{Y}}\!+\!(1\!+\!\omega \check{Y})\gamma^{2})\nonumber\\
		\times&q^{-\frac{1}{2}}\tilde{K}^{[i}_{a}\tilde{K}^{j]}_{b}E^{a}_{i}E^{b}_{j},\nonumber\\
		C_{3}[N]=&\int_{\Sigma}{d^3x}N\frac{2/3}{16\pi G(1+\omega \check{Y})}q^{-\frac{1}{2}}(\tilde{K}^{i}_{a}E^{a}_{i})^{2},\nonumber\\
		C_{4}[N]=&\int_{\Sigma}{d^3x}N\frac{2}{3\omega}q^{-\frac{1}{2}}\tilde{K}^{i}_{a}E^{a}_{i}\pi^{Y},\nonumber\\
		C_{5}[N]=&\int_{\Sigma}{d^3x}N\frac{8\pi G}{3\omega^{2}}(1+\omega \check{Y})q^{-\frac{1}{2}}(\pi^{Y})^{2},\nonumber\\
		C_{6}[N]=&\int_{\Sigma}{d^3x}N\frac{\omega}{8\pi G}q^{\frac{1}{2}}D_{a}D^{a}\check{Y},\nonumber\\
		C_{7}[N]=&-\int_{\Sigma}{d^3x}\frac{Nq^{-\frac{1}{2}}}{8\rho}\pi_{a}\pi^{a},\nonumber\\
		C_{8}[N]=&-\int_{\Sigma}{d^3x}N\rho q^{\frac{1}{2}}\Omega_{ab}\Omega^{ab},\nonumber\\
		C_{9}[N]\!=\!&\int_{\Sigma}\!{d^3x}Nq^{-\frac{1}{2}}(E^{ai}Y_{a}E^{b}_{j}Y_{b}\!-\!q\check{Y})^{1/2}D_{c}\pi^{c}.\label{nine terms}
\end{align}
To define their corresponding operators, we only need to specify their actions on the basis vector $T_{s,u,c}$. For simplicity, we adjust the basis vector $T_{s,u,c}$ to an adapted form before being acted, such that: (i) all vertices of graph $\alpha$ except the pseudo-vertices are the beginning points of edges $e\in E(\alpha)$ with them as endpoints; (ii) all intersections of $V(\alpha)$ with the graph $\beta$ are vertices of $\beta$, and they serve as the beginning points of edges $e\in E(\beta)$ with them as endpoints. By introducing the characteristic function $\chi_{\delta}(x,y)$ such that $\lim_{\delta\to0}\chi_{\delta}(x,y)/\delta^3=\delta^{(3)}(x,y)$, and using the point-splitting scheme \cite{Thiemann4,Thiemann:1997rt,Xiangdong1,Zhang:2011qq,Xiangdong2,Zhang:2011gn,Jinsong,Lewandowski1,Alesci:2015wla}, the expressions in (\ref{nine terms}) can be regularized respectively as:
\begin{align}
	C_{1}[N]\!&=\!\lim\limits_{\delta\to0}\int_{\Sigma}\!{d^3x}\!N\!(x)\!V^{-\frac{1}{2}}_{(x,\delta)}\!\frac{1\!+\!\omega \check{Y}\!(x)}{16\pi G}\!\epsilon_{i}^{\,j\!k}\!F^{i}_{ab}\!(x)\nonumber\\
	&\times\!E^{a}_{j}(x)\!\int_{\Sigma}\!{d^3y}\!\chi_{\delta}(x,y)\!E^{b}_{k}(y)V^{-\frac{1}{2}}_{(y,\delta)},\label{SC1}\\
	\nonumber\\
	C_{2}[N]&=\lim\limits_{\delta\to0}\int_{\Sigma}{d^3x}\frac{-N(x)}{8\pi G}V^{-\frac{1}{2}}_{(x,\delta)}(\dfrac{1}{1+\omega \check{Y}(x)}\nonumber\\
	&+(1+\omega \check{Y}(x))\gamma^{2})\tilde{K}^{[i}_{a}\tilde{K}^{j]}_{b}(x)E^{a}_{i}(x)\nonumber\\
	&\times\int_{\Sigma}{d^3y}\chi_{\delta}(x,y)E^{b}_{j}(y)V^{-\frac{1}{2}}_{(y,\delta)},\label{SC2}\\
	\nonumber\\
	C_{3}[N]\!&=\!\lim\limits_{\delta\to0}\!\int_{\Sigma}\!{d^3x}\!N(x)\!V^{-\frac{1}{2}}_{(x,\delta)}\!\frac{(2/3)\!\tilde{K}^{i}_{a}(x)\!\tilde{K}^{j}_{b}(x)}{16\pi G(1\!+\!\omega \check{Y}(x))}\nonumber\\
	&\times\!E^{a}_{i}(x)\!\int_{\Sigma}\!{d^3y}\!\chi_{\delta}(x,y)E^{b}_{j}(y)V^{-\frac{1}{2}}_{(y,\delta)},\label{SC3}\\
	\nonumber\\
	C_{4}[N]&=\lim\limits_{\delta\to0}\int_{\Sigma}{d^3x}N(x)\frac{2}{3\omega}V^{-\frac{1}{2}}_{(x,\delta)}\tilde{K}^{i}_{a}(x)E^{a}_{i}(x)\nonumber\\
	&\times\int_{\Sigma}{d^3y}\chi_{\delta}(x,y)\pi^{Y}(y)V^{-\frac{1}{2}}_{(y,\delta)},\label{SC4}\\
	\nonumber\\
	C_{5}[N]&=\lim\limits_{\delta\to0}\int_{\Sigma}{d^3x}N(x)\frac{8\pi G}{3\omega^{2}}V^{-\frac{1}{2}}_{(x,\delta)}(1\!+\!\omega \check{Y}(x))\nonumber\\
	&\times\!\pi^{Y}(x)\!\int_{\Sigma}\!{d^3y}\!\chi_{\delta}(x,y)\!\pi^{Y}(y)\!V^{-\frac{1}{2}}_{(y,\delta)},\label{SC5}\\
	\nonumber\\
	C_{6}[N]&=\lim\limits_{\delta\to0}\int_{\Sigma}{d^3x}N(x)\frac{\omega}{8\pi{G}}V^{-\frac{1}{2}}_{(x,\delta)}q^{1/2}(x)\nonumber\\
	&\times\int_{\Sigma}{d^3y}\chi_{\delta}(x,y)V^{-\frac{1}{2}}_{(y,\delta)}q^{1/2}(y)D_{a}D^{a}\check{Y}(y)\nonumber\\
	&=\lim\limits_{\delta\to0}-\int_{\Sigma}{d^3x}N(x)\frac{\omega}{8\pi{G}}V^{-\frac{1}{2}}_{(x,\delta)}q^{1/2}(x)\nonumber\\
	&\times\!\int_{\Sigma}\!{d^3y}\!D_{a}(\chi_{\delta}(x,y)V^{-\frac{1}{2}}_{(y,\delta)})q^{1/2}(y)D^{a}\check{Y}(y)\nonumber\\
	&=\lim\limits_{\delta\to0}-\int_{\Sigma}{d^3x}N(x)\frac{\omega}{8\pi{G}}V^{-\frac{1}{2}}_{(x,\delta)}q^{1/2}(x)\nonumber\\
	&\times\!\int_{\Sigma}\!{d^3y}\!D_{a}(\chi_{\delta}(x,y)V^{-\frac{1}{2}}_{(y,\delta)})\!(D_{b}\check{Y}\!(y))\!E^{ai}(y)\nonumber\\
	&\times\!V^{-\frac{1}{2}}_{(y,\delta)}\!\int_{\Sigma}\!{d^3z}\!\chi_{\delta}(y,z)E^{b}_{i}(z)\!V^{-\frac{1}{2}}_{(z,\delta)},\label{SC6}\\
	\nonumber\\
	C_{7}[N]&=\lim\limits_{\delta\to0}-\int_{\Sigma}{d^3x}\frac{N(x)}{32\rho}V^{-3/2}_{(x,\delta)}\delta_{kl}\epsilon^{ijk}\epsilon_{eab}\nonumber\\
	&\times\!\pi^{e}(x)E^{a}_{i}(x)\!\int_{\Sigma}{d^3y}\chi_{\delta}(x,y)E^{b}_{j}(y)\nonumber\\
	&\times\!\int_{\Sigma}{d^3z}\chi_{\delta}(y,z)\epsilon^{mnl}\epsilon_{fcd}\pi^{f}(z)E^{c}_{m}(z)\nonumber\\
	&\times\!\int_{\Sigma}{d^3w}\chi_{\delta}(z,w)E^{d}_{n}(w)V^{-3/2}_{(w,\delta)},\label{SC7}\\
	\nonumber\\
	C_{8}[N]&=\lim\limits_{\delta\to0}-\int_{\Sigma}{d^3x}N(x)\frac{\rho}{2}{V}^{-3/2}_{(x,\delta)}\delta_{kl}\epsilon^{ijk}\nonumber\\
	&\times\!\Omega_{ab}(x)E^{a}_{i}(x)\!\int_{\Sigma}{d^3y}\chi_{\delta}(x,y)E^{b}_{j}(y)\nonumber\\
	&\times\!\int_{\Sigma}{d^3z}\chi_{\delta}(y,z)\epsilon^{mnl}\Omega_{cd}(z)E^{c}_{m}(z)\nonumber\\
	&\times\!\int_{\Sigma}{d^3w}\chi_{\delta}(z,w)E^{d}_{n}(w)V^{-3/2}_{(w,\delta)},\label{SC8}\\
	\nonumber\\
	C_{9}[N]&=\lim\limits_{\delta\to0}\int_{\Sigma}{d^3x}N(x)V^{-\frac{1}{2}}_{(x,\delta)}(E^{ai}Y_{a}E^{b}_{j}Y_{b}\nonumber\\
	&-q\check{Y})^{\frac{1}{2}}(x)\!\int_{\Sigma}\!{d^3y}\!\chi_{\delta}(x,y)\!V^{-\frac{1}{2}}_{(y,\delta)}\!D_{c}\pi^{c}(y)\nonumber\\
	&=-\lim\limits_{\delta\to0}\int_{\Sigma}{d^3x}N(x)V^{-\frac{1}{2}}_{(x,\delta)}(E^{ai}Y_{a}E^{b}_{j}Y_{b}\nonumber\\
	&-q\check{Y})^{\frac{1}{2}}(x)\!\int_{\Sigma}\!{d^3y}\!D_{c}\!(\chi_{\delta}(x,y)\!V^{-\frac{1}{2}}_{(y,\delta)})\pi^{c}(y),\label{SC9}
\end{align}
where $V_{(x,\delta)}\equiv\int_{\Sigma}{d^3y}\chi_{\delta}(x,y)\sqrt{q}(y)$. For the expressions of (\ref{SC1}), (\ref{SC2}), and (\ref{SC3}), we replace $E^{a}_{i}(x)$ with $-i\hbar8\pi{G}\gamma\delta/\delta{A^{i}_{a}(x)}$ and apply them to $T_{s,u,c}$. Similar to the results in Refs.\cite{Jinsong,Alesci:2015wla,Thiemann:1997rt}, the results become sums over integrals for all edges $e\in E(\alpha)$. By decomposing each edge into $\mathrm{N}$ segments of parameter length $\epsilon$ along the domain $[0,1]$, the integral along the edge is approximated by a Riemann sum. By replacing $V^{-1/2}_{(x,\delta)}$ in the summation with $\hat{V}^{-1/2}_{(x,\delta)}\equiv\lim\limits_{\varsigma\to0}(\hat{V}_{(x,\delta)}+\varsigma)^{-1}\hat{V}^{1/2}_{(x,\delta)}$ and letting $\delta\to0$, it follows that, due to the vanishing of the operator $\hat{V}^{-1/2}_{x}\equiv\lim\limits_{\delta\to0}\hat{V}^{-1/2}_{(x,\delta)}$ at coplanar vertices, only segments around the non-coplanar vertices $v\in V(\alpha)$ contribute nontrivial results. By combining these segments with $F^{i}_{ab}$ and $\tilde{K}^{i}_{a}$, we obtain \cite{Jinsong,Alesci:2015wla,Thiemann4}:
\begin{align}
	\epsilon^2F^{i}_{ab}(v)\dot{e}^{a}(v)\dot{e}'^{b}(v)\approx\frac{6}{W_{l}^{2}}\varepsilon(e,e')Tr(h_{\alpha^{\epsilon}_{e,e'}}^{(l)}\tau^{(l)i}),\label{approx1}
\end{align}	
\begin{align}
	\epsilon\tilde{K}^{i}_{a}(v)\dot{e}^{a}(v)\!&\approx\!\frac{3}{8G\pi\gamma^3W_{l}^{2}}T\!r(\tau^{(l)i}(h_{s(e)}^{(l)})^{\!-\!1}\nonumber\\
		&\times\!\{h_{s(e)}^{(l)},\{C_{1}[1]|_{\omega=0},V\}\}),\label{approx2}
\end{align}
where, $V\equiv\int_{\Sigma}{d^3x}\sqrt{q}(x)$, $(\tau^{(l)i})^{m}_{n}\equiv{\left<lm\right|}\tau^{i}{\left|ln\right>}$, $W_{l}^{2}\equiv\!l(l+1)(2l+1)$, $\varepsilon(e,e')=0$ if $\dot{e}^{a}(v)$ is parallel to $\dot{e}'^{b}(v)$ and otherwise $\varepsilon(e,e')=1$, $s(e)$ is a small segment of $e$ from the vertex $v$ with $\epsilon$ as the parameter length, $\alpha^{\epsilon}_{e,e'}$ is the loop formed by adding an arc from the final point of $s(e)$ to the final point of $s(e')$. Note that the assignment of $\alpha^{\epsilon}_{e,e'}$ is diffeormorphism covariant in the following sense \cite{Thiemann,Thiemann4}: (i) $\forall\,\epsilon,\epsilon'$, $\exists\,\varphi_{\alpha}\in{Diff}$ such that $\varphi_{\alpha}(\alpha)=\alpha$ and $\varphi_{\alpha}(\alpha_{e,e'}^{\epsilon'})=\alpha_{e,e'}^{\epsilon}$; (ii) $\forall\,\varphi\in{Diff}$, $\exists\,\varphi'\in{Diff}$ such that $\varphi'(\varphi(\alpha))=\varphi(\alpha)$ and $\varphi'(\alpha^{\epsilon}_{\varphi(e),\varphi(e')})=\varphi(\alpha^{\epsilon}_{e,e'})$. By replacing the regularized variables with the corresponding operators and the Poisson bracket with the commutator multiplying by $(i\hbar)^{-1}$, we obtain the operator $\hat{K}^{i}_{(v,e)}$ corresponding to the right side of (\ref{approx2}). Since the terms involving the scalar field $\check{Y}$ also contribute nontrivial result at the vertex $v$, we introduce a small constant $c_{0}$ to regularize them as follows \cite{Xiangdong1,Zhang:2011qq,Xiangdong2,Zhang:2011gn}:
\begin{align}
\check{Y}(v)\approx\frac{1}{2ic_{0}}(\exp(ic_{0}\check{Y}(x))-\exp(-ic_{0}\check{Y}(x))),\label{Y1}
\end{align}
\begin{align}
\frac{1}{1\!+\!\omega \check{Y}(v)}\!&=\!(1\!+\!\omega \check{Y}(v))\frac{16}{\omega^4}\nonumber\\
&\times\!\Big(\{\left|1\!+\!\omega \check{Y}(v)\right|^{1/2},\!\int_{\Sigma}d^3x\pi^Y\}\Big)^{4}.\label{Y2}
\end{align}
The expressions in (\ref{Y1}) and (\ref{Y2}) already have clear quantum analogs and hence can be promoted to the corresponding operators. Collecting the above results, the operators corresponding to $C_{1}[N]$, $C_{2}[N]$ and $C_{3}[N]$ act on the basis as:
\begin{align}
	\hat{C}_{1}[N]\!\cdot\! T_{s,u,c}\!&=\!\frac{(8G\pi\gamma)^{2}6}{16G\pi{W_{l}^{2}}}\!\sum\limits_{v\in{V(\alpha)}}\!N\!(v)\!\hat{V}^{-\frac{1}{2}}_{v}(1\!+\!\omega\hat{\check{Y}}\!(v)\!)\nonumber\\
	&\times\!\sum\limits_{\left.e\right|_{v},\left.e'\right|_{v}\in{E(\alpha)}}\!\varepsilon(e,e')\epsilon_{i}^{\,j\!k}\!T\!r(h_{\alpha^{\epsilon}_{e,e'}}^{(l)}\tau^{(l)i})\nonumber\\
	&\times\hat{J}^{(v,e)}_{j}\hat{J}^{(v,e')}_{k}\hat{V}^{-\frac{1}{2}}_{v}\!\cdot\!T_{s,u,c},\\
	\hat{C}_{2}[N]\!\cdot\! T_{s,u,c}\!&=\!-\frac{(8G\pi\gamma)^{2}}{8G\pi}\!\sum\limits_{v\in{V(\alpha)}}\!N\!(v)\!\hat{V}^{-\frac{1}{2}}_{v}\!(\widehat{\dfrac{1}{1\!+\!\omega\check{Y}}}\nonumber\\
	&+\!(1\!+\!\omega\hat{\check{Y}}(v))\gamma^{2})\!\sum\limits_{\left.e\right|_{v},\left.e'\right|_{v}\in{E(\alpha)}}\!\varepsilon(e,e')\nonumber\\
	&\times\!\hat{K}^{[i}_{(v,e)}\!\hat{K}^{j]}_{(v,e')}\!\hat{J}^{(v,e)}_{i}\!\hat{J}^{(v,e')}_{j}\!\hat{V}^{-\frac{1}{2}}_{v}\!\cdot\!T_{s,u,c},\\
	\hat{C}_{3}[N]\!\cdot\! T_{s,u,c}\!&=\!\frac{2/3(8G\pi\gamma)^{2}}{16G\pi}\!\sum\limits_{v\in{V(\alpha)}}\!N\!(v)\!\hat{V}^{-\frac{1}{2}}_{v}\!\widehat{\dfrac{1}{1\!+\!\omega \check{Y}}}\nonumber\\
	&\times\!\sum\limits_{\left.e\right|_{v},\left.e'\right|_{v}\in{E(\alpha)}}\!\hat{K}^{i}_{(v,e)}\!\hat{K}^{j}_{(v,e')}\!\hat{J}^{(v,e)}_{i}\!\hat{J}^{(v,e')}_{j}\nonumber\\
	&\times\!\hat{V}^{-\frac{1}{2}}_{v}\!\cdot\!T_{s,u,c},
\end{align}
where $\hat{J}^{(v,e)}_{i}\!\cdot\!T_{s,u,c}\!\equiv\!i\hbar\partial_{t}T_{s}(\!\cdots,h_{e}e^{t\tau_{i}},\cdots)\!|_{t=0}\\\times\!T_{u}T_{c}$, and $l$ is chosen as an appropriate half-integer such that the operator does not change the valence of any vertices in $\alpha$, ensuring that its adjoint operator is densely defined.

By noticing the structural similarity between the expressions of (\ref{SC6}), (\ref{SC7}), (\ref{SC8}), and that of (\ref{SC1}), following the above discussions, it is sufficient to regularize the following expressions around the vertex $v\in V(\alpha)$ by employing techniques analogous to those in Refs.\cite{Xiangdong1,Zhang:2011qq,Xiangdong2,Zhang:2011gn,Thiemann:1997rt}:
\begin{align}
&\epsilon\dot{e}'^{a}(v)D_{a}(\chi_{\delta}(\!x,\!v)\!V^{-1/2}_{(v,\delta)})\nonumber\\
\approx&\chi_{\delta}(\!x,\!f(s(e')))V^{-1/2}_{(f(\!s(e')\!),\delta)}\!-\!\chi_{\delta}(\!x,\!v)V^{-1/2}_{(v,\delta)},
\end{align}
\begin{align}
	&\epsilon\dot{e}^{b}\!(v)\!D_{b}\check{Y}\!(v)\!\approx\!\frac{\exp(ic_{0}(\check{Y}\!(f\!(s\!(e)))\!-\!\check{Y}\!(v)))\!-\!1}{ic_{0}},\\
	&\epsilon^2\dot{e}^{c}(v)\dot{e}'^{d}(v)\epsilon_{fcd}\pi^{f}(v)\approx2\pi[S^{\epsilon}_{e,e'}],\\
	&\epsilon^2\dot{e}^{c}(v)\dot{e}'^{d}(v)\Omega_{cd}(v)\!\approx\!2\frac{\exp(i\zeta_{0}\int_{\alpha^{\epsilon}_{e,e'}}Y_{a})\!-\!1}{i\zeta_{0}},
\end{align}
where $S^{\epsilon}_{e,e'}$ is a surface enclosed by $\alpha^{\epsilon}_{e,e'}$. Combining with (\ref{flux}), we can promote $C_{6}[N]$, $C_{7}[N]$ and $C_{8}[N]$ to the corresponding operators acting on the basis as
\begin{align}
	\hat{C}_{6}[N]\!\cdot\! T_{s,u,c}&\!=\!\frac{\omega(8G\pi\gamma)^{2}}{8\pi{G}}\!\sum\limits_{v\in{V(\alpha)}}\!\sum\limits_{\left.e\right|_{v},\left.e'\right|_{v}\in{E(\alpha)}}\!N\!(v)\nonumber\\
	&\times\!\frac{\exp(ic_{0}(\check{Y}(\!f(s(e))\!)\!-\!\check{Y}(v)))\!-\!1}{ic_{0}}\nonumber\\
	&\times\!\hat{V}^{-\frac{1}{2}}_{v}\hat{J}^{(v,e')i}\hat{J}^{(v,e)}_{i}\hat{V}^{-\frac{1}{2}}_{v}\!\cdot\!T_{s,u,c},\\
	\hat{C}_{7}[N]\!\cdot\! T_{s,u,c}\!&=\!-\frac{(8G\pi\gamma)^{4}}{8\rho}\!\sum\limits_{v\in{V(\alpha)}\cap{V(\beta)}}\!N\!(v)\!\hat{V}^{-\frac{3}{2}}_{v}\nonumber\\
	&\times\!\delta_{ij}\hat{E}^{i}_{v}\hat{E}^{j}_{v}\hat{V}^{-\frac{3}{2}}_{v}\!\cdot\!T_{s,u,c},\\
	\hat{C}_{8}[N]\!\cdot\! T_{s,u,c}\!&=\!-2\rho(8G\pi\gamma)^{4}\!\sum\limits_{v\in{V(\alpha)}}\!N(v)\hat{V}^{-\frac{3}{2}}_{v}\nonumber\\
	&\times\!\delta_{ij}\hat{B}^{i}_{v}\hat{B}^{j}_{v}\hat{V}^{-\frac{3}{2}}_{v}\!\cdot\!T_{s,u,c},
\end{align}
with
\begin{align}
	\hat{E}^{i}_{v}\!\cdot\! T_{s,u,c}\!&:=\!\sum\limits_{\left.e\right|_{v},\left.e'\right|_{v}\in{E(\alpha)},\left.e''\right|_{v}\in{E(\beta)}}\varepsilon(\!e,\!e'\!,e''\!)\epsilon^{ijk}\nonumber\\
	&\times\!\hat{J}^{(v,e')}_{j}\hat{J}^{(v,e'')}_{k}\frac{\hbar}{2}\zeta_{e''}\!\cdot\!T_{s,u,c},\\
	\hat{B}^{i}_{v}\!\cdot\! T_{s,u,c}\!&:=\!\sum\limits_{\left.e\right|_{v},\left.e'\right|_{v}\in{E(\alpha)}}\!\varepsilon(\!e,\!e'\!)\!\frac{\exp(i\zeta_{0}\!\int_{\alpha^{\epsilon}_{e,e'}}\!Y_{a})\!-\!1}{i\zeta_{0}}\nonumber\\
	&\times\!\epsilon^{ijk}\!\hat{J}^{(v,e)}_{j}\hat{J}^{(v,e')}_{k}\!\cdot\!T_{s,u,c},
\end{align}
where $\varepsilon(e,e',e'')\equiv\text{sgn}(\epsilon_{abc}\dot{e}^{a}(v)\dot{e}'^{b}(v)\dot{e}''^{c}(v))$, and $\zeta_{0}$ is selected as an appropriate real number so that the operator does not decrease the valence of any vertices in $\beta$, thereby ensuring its adjoint operator is densely defined.

For the expressions of (\ref{SC4}) and (\ref{SC5}), by replacing $\pi^{Y}(x)$ with $-i\hbar\delta/\delta{\check{Y}(x)}$, $E^{a}_{i}(x)$ with $-i\hbar8\pi{G}\gamma\delta/\delta{A^{i}_{a}(x)}$, we obtain their corresponding operators acting on the basis as:
\begin{align}
	\hat{C}_{4}[N]\!\cdot\! T_{s,u,c}\!&=\!\lim\limits_{\delta\to0}\!\sum\limits_{v\in{X}}\!\sum\limits_{e\in{E(\alpha)}}\!\int_{0}^{1}\!dt\![\frac{2\!N\!(e(t))}{3\omega}\!V^{-\frac{1}{2}}_{(e(t),\delta)}\nonumber\\
	&\times\!\tilde{K}^{i}_{a}(e(t))\dot{e}^{a}(t)i\hbar8G\pi\gamma{X_{e,i}(t)}\nonumber\\
	&\times\!\chi_{\delta}(e(t),v)]\!\hbar c_{v}V^{-\frac{1}{2}}_{(v,\delta)}\!\cdot\!T_{s,u,c},\label{RC4}\\
	\hat{C}_{5}[N]\!\cdot\! T_{s,u,c}\!&=\!\lim\limits_{\delta\to0}\sum\limits_{v\in{X}}\sum\limits_{v'\in{X}}N(v')\frac{8\pi G}{3\omega^{2}}V^{-\frac{1}{2}}_{(v',\delta)}\nonumber\\
	&\times\!(1\!+\!\omega \check{Y}\!(v'))\hbar c_{v'}\chi_{\delta}(v',v)\nonumber\\
	&\times\!\hbar c_{v}V^{-\frac{1}{2}}_{(v,\delta)}\!\cdot\!T_{s,u,c},\label{RC5}
\end{align}
where $X_{e,i}(t)\equiv\!Tr(h_{e[1,s]}\tau_{i}h_{e[s,0]}\partial/\partial{h_{e}})$. By inserting $\mathrm{N}-1$ points $0=t_{0},t_{1},\cdots,t_{\mathrm{N}}=1$ to divide the domain $[0,1]$ into $\mathrm{N}$ intervals, and setting $t_{n}-t_{n-1}=\epsilon$, the integral in (\ref{RC4}) can be expressed as a Riemann sum and thus
\begin{align}
		\hat{C}_{4}[N]\!\cdot\! T_{s,u,c}\!&=\!\lim\limits_{\epsilon,\delta\to0}\!\sum\limits_{v\in{X}}\!\sum\limits_{e\in{E(\alpha)}}\!\sum\limits_{n=1}^{\mathrm{N}}[\frac{2N(e(t_{n-1}))}{3\omega}\nonumber\\
		&\times\!V^{-\frac{1}{2}}_{(e(t_{n-1}),\delta)}\tilde{K}^{i}_{a}(e(t_{n-1}))\epsilon\dot{e}^{a}(t_{n-1})\nonumber\\
		&\times\!i\hbar8G\pi\gamma{X_{e,i}(t_{n-1})}\chi_{\delta}(e(t_{n-1}),v)]\nonumber\\
		&\times\hbar c_{v}V^{-\frac{1}{2}}_{(v,\delta)}\cdot T_{s,u,c}.\label{RRC4}
\end{align}
Replacing the variables in (\ref{RRC4}) and (\ref{RC5}) with the corresponding operators, and taking $\delta\to0$, the operators corresponding to $C_{4}[N]$ and $C_{5}[N]$ act on the basis as
\begin{align}
	\hat{C}_{4}[N]\!\cdot\! T_{s,u,c}\!&=\!\sum\limits_{v\in{X}\cap{V(\alpha)}}\sum\limits_{\left.e\right|_{v}\in{E(\alpha)}}N(v))\frac{2}{3\omega}\hat{V}^{-\frac{1}{2}}_{v}\nonumber\\
	&\times\!\hat{K}^{i}_{(v,e)}\!8G\pi\gamma\hat{J}^{(v,e)}_{i}\hbar c_{v}\hat{V}^{-\frac{1}{2}}_{v}\!\cdot\!T_{s,u,c},\\
	\hat{C}_{5}[N]\!\cdot\!T_{s,u,c}\!&=\!\sum\limits_{v\in{X}\cap{V(\alpha)}}\!N\!(v)\!\frac{8\pi G}{3\omega^{2}}\!\hat{V}^{-\frac{1}{2}}_{v}\!(1\!+\!\omega \hat{\check{Y}}\!(v))\nonumber\\
	&\times\!(\hbar c_{v})^{2}\hat{V}^{-\frac{1}{2}}_{v}\!\cdot\! T_{s,u,c}.
\end{align}

By substituting $\pi^{c}(y)$ with $-i\hbar\delta/\delta{Y_{c}(y)}$ in the expression of (\ref{SC9}), we obtain the corresponding operator acting on the basis as
\begin{align}
		\hat{C}_{9}[N]\!\cdot\! T_{s,u,c}\!&=\!-\lim\limits_{\delta\to0}\!\int_{\Sigma}\!{d^3x}\!N(x)\!V^{-1/2}_{(x,\delta)}\!(E^{ai}\!Y_{a}\!E^{b}_{j}\!Y_{b}\nonumber\\
		&\!-\!q\check{Y})^{\frac{1}{2}}(x)\!\sum_{v\in V(\beta)}\!\chi_{\delta}(x,v)V^{-\frac{1}{2}}_{(v,\delta)}\nonumber\\
		&\times\!\sum_{\left.e\right|_{v}\in{E(\beta)}}
		\!(\delta_{v,f(e)}\!-\!\delta_{v,b(e)})\hbar\zeta_{e}\!\cdot\!T_{s,u,c}.\label{RRC9}
\end{align}
Analogous to the situation in Refs.\cite{Alesci:2015wla,Lewandowski2}, by introducing a partition $\Sigma=\cup_{r}\Sigma^{\iota}_{r}$ parameterized by $\iota$, such that all cells $\Sigma^{\iota}_{r}$ uniformly shrink as $\iota\to0$, we obtain the following regularized expression:
\begin{align}
	&\int_{\Sigma}\!{d^3x}\!N(x)\!\chi_{\delta}(x,v)V^{-\frac{1}{2}}_{(x,\delta)}\!(E^{ai}\!Y_{a}\!E^{b}_{j}\!Y_{b}\!-\!q\check{Y})^{\frac{1}{2}}\!(x)\nonumber\\
	=&\lim_{\iota\to0}\sum\limits_{r}\text{sgn}(N(x_{r}))\sqrt{A_{\iota}^{2}+B_{\iota}^{2}},\label{RC9}
\end{align}
where
\begin{align}
	A_{\iota}\!:=\!&\int_{\Sigma^{\iota}_{r}}\!d^3x\!N(x)\!\chi_{\delta}(x,v)\!V^{-\frac{1}{2}}_{(x,\delta)}\!\sqrt{E^{ai}Y_{a}E^{b}_{j}Y_{b}}(x),\\
	B_{\iota}\!:=\!&\int_{\Sigma^{\iota}_{r}}\!d^3x\!N(x)\!\chi_{\delta}(x,v)\!V^{-\frac{1}{2}}_{(x,\delta)}|\check{Y}|^{1/2}(x)q^{\frac{1}{2}}(x).
\end{align}
As shown in Ref.\cite{Yongge1}, $A_{\iota}$ can be promoted to the following operator acting on the edges as
\begin{align}
	\hat{A}_{\iota}\!\cdot\!T_{s,u,c}\!&=\!8G\pi\gamma\hbar\sum\limits_{e\in{E(\alpha)}}\!\int_{e\cap\Sigma^{\iota}_{r}}\!dt|\dot{e}^{a}(t)\!N(e(t))\nonumber\\
	&\times\!\chi_{\delta}(e(t),v)V^{-\frac{1}{2}}_{(e(t),\delta)}Y_{a}(e(t))|\nonumber\\
	&\times\!\sqrt{j_{e}(j_{e}+1)}\,T_{s,u,c}.
\end{align}
Since $B_{\iota}$ is similar to a smeared volume, it can be promoted to the following operator acting on the vertices as
\begin{align}
	\hat{B}_{\iota}\!\cdot \!T_{s,u,c}\!&=\!\sum\limits_{v'\in{V(\alpha)\cap\Sigma^{\iota}_{r}}}\!N(v')\chi_{\delta}(v',v)|\hat{\check{Y}}(v')|^{\frac{1}{2}}\nonumber\\
	&\times\!{V}^{-\frac{1}{2}}_{(v',\delta)}\hat{V}_{v'}\!\cdot\!T_{s,u,c}.
\end{align}
Similar to the case discussed in Ref.\cite{Lewandowski2}, for sufficiently small $\iota$, if we replace $A_{\iota}$ and $B_{\iota}$ in (\ref{RC9}) with their corresponding operators, then when $\Sigma^{\iota}_{r}$ does not contain a vertex, only $\hat{A}_{\iota}$ contributes, and when $\Sigma^{\iota}_{r}$ contains a vertex, only $B_{\iota}$ contributes. Therefore, (\ref{RRC9}) can be rewritten as 
\begin{align}
\hat{C}_{9}[N]\!\cdot\!T_{s,u,c}&\!=\!\lim_{\delta\to0}\!\sum_{v\in V(\beta)}\![8G\pi\gamma\hbar\!\sum\limits_{e\in{E(\alpha)}}\!\int_{e}\!dt\!|\dot{e}^{a}(t)\nonumber\\
&\times\!Y_{a}(e(t))|N(e(t))\chi_{\delta}(e(t),v)V^{-\frac{1}{2}}_{(e(t),\delta)}\nonumber\\
&\times\!\sqrt{j_{e}(j_{e}\!+\!1)}\!+\!\sum_{v'\in{V(\alpha)}}\!N(v')\!\chi_{\delta}(v',v)\nonumber\\
&\times\!|\check{Y}(v')|^{\frac{1}{2}}{V}^{-\frac{1}{2}}_{(v',\delta)}\hat{V}_{v'}]V^{-\frac{1}{2}}_{(v,\delta)}\nonumber\\
&\times\!\sum_{\left.e\right|_{v}\in{E(\beta)}}\!(\delta_{v,b(e)}\!-\!\delta_{v,f(e)})\hbar\zeta_{e}\!\cdot\! T_{s,u,c}.\label{RRRC9}
\end{align}
By decomposing the edge integral in (\ref{RRRC9}) into a Riemann sum of $\mathrm{N}$ segments, each of length $\epsilon$, and replacing $V^{-1/2}_{(x,\delta)}$ with $\hat{V}^{-1/2}_{(x,\delta)}$, it is easy to see that only the segments near the vertices $v\in V(\alpha)$ contribute nontrivial results. Hence we introduce the following regularization:
\begin{align}
	\epsilon\!\left|\!\dot{e}^{a}\!(v)\!Y_{a}\!(v)\!\right|\!\approx&\!\left|\! \frac{\exp(i\zeta_{0}\!\int_{s(e)}\!Y)\!-\!\exp(-i\zeta_{0}\!\int_{s(e)}\!Y)}{2i\zeta_{0}}\!\right|\nonumber\\
	=:&|\hat{\Upsilon}_{s(e)}|.
\end{align} 
Collecting the above results, the operator corresponding to $C_{9}[N]$ acts on the basis as 
\begin{align}
		\hat{C}_{9}[N]\!\cdot\! T_{s,u,c}\!&=\!\sum_{v\in\!{V(\alpha)}\cap{V(\beta)}}\!N(v)\!\hat{V}^{-\frac{1}{2}}_{v}[8G\pi\gamma\hbar\nonumber\\
		&\times\!\sum\limits_{\left.e\right|_{v}\in{E(\alpha)}}\!|\!\hat{\Upsilon}_{s(e)}\!|\!\sqrt{\!j_{e}(j_{e}\!+\!1)}\!\nonumber\\
		&+\!|\hat{\check{Y}}(v)|^{\frac{1}{2}}\hat{V}_{v}]\!\sum_{\left.e'\right|_{v}\in{E(\beta)}}\!(\delta_{v,b(e)}\!-\!\delta_{v,f(e)})\nonumber\\
		&\times\!\hbar\zeta_{e'}\hat{V}^{-\frac{1}{2}}_{v}\!\cdot\!T_{s,u,c}.
\end{align}

\subsection{Vertex Hilbert Space}
Since the assignments of $s(e)$ and $\alpha^{\epsilon}_{e,e'}$ are diffeomorphism covariant, we can promote $\hat{C}^{\epsilon}[N]\equiv\sum_{m=1}^{9}\hat{C}_{m}[N]$ to an operator on the vertex Hilbert space introduced in Refs.\cite{Jinsong,Lewandowski1,Alesci:2015wla} to remove the regularization parameter $\epsilon$. Note that the operator $\hat{C}^{\epsilon}[N]$ acts only on the non-coplanar vertices with valence greater than 3 in graph $\alpha$, denoted as $V_{np4}(\alpha)$. We define the anti-linear rigging map $\tilde{\eta}:\mathcal{D}^{G}_{kin}\to(\mathcal{D}^{G}_{kin})^{*}$ as 
\begin{equation}
	\tilde{\eta}(T_{s,u,c})\!:=\!\sum_{\varphi\in{D\!i\!f\!f_{np4}/T\!D\!i\!f\!f_{\alpha,\beta,X}}} \!\big<\hat{U}(\varphi)\!\cdot\!T_{s,u,c},\cdot\big>_{kin},
\end{equation}
where $Diff_{np4}\!:=\!\{\varphi\in{Diff}\mid\varphi(v)\!=\!v,\forall v\in\!V_{np4}(\alpha)\}$. Then, the inner product on $\tilde{\eta}[\mathcal{D}^{G}_{kin}]$ can be defined by
\begin{equation}\label{scalar product5}
	\left<\tilde{\eta}(T_{s,u,c}),\tilde{\eta}(T_{s',u',c'})\right>_{vtx}=[\tilde{\eta}(T_{s,u,c})](T_{s',u',c'}).
\end{equation}
By completing $\tilde{\eta}[\mathcal{D}^{G}_{kin}]$ using (\ref{scalar product5}), one obtains the vertex Hilbert space $\mathcal{H}_{vtx}$. 

It turns out that $\hat{C}^{\epsilon}[N]$ can be promoted to an $\epsilon$-independent operator on $\mathcal{H}_{vtx}$ through the following dual action:
\begin{equation}
	[\hat{C}^{*}\![N]\!\tilde{\eta}(T_{s,u,c})]\!\cdot\! T_{s',u',c'}\!:=\!\tilde{\eta}(T_{s,u,c})\!(\hat{C}^{\epsilon}\![N]\!\cdot\!T_{s',u',c'}).
\end{equation}
The constraint operator $\hat{C}^{*}[N]$ has several desirable properties, such as: its adjoint operator being densely defined, the quantum constraint algebra being anomaly-free, and the group averaging of the kernel of $\hat{C}^{*}[N]$ with respect to the residual diffeomorphism $Diff/Diff_{np4}$ yielding the physical Hilbert space with all constraints resolved.

\section{Deparametrization in Spherically Symmetric Model}\label{DVTGSSM}

\subsection{A Review on Deparametrization}\label{Deparametrization}
Let $\Gamma_{pre}$ be a phase space consisting of canonical variables $(T^{I}(x),P_{J}(x'))$, $(q_{ab}(x),p^{cd}(x'))$, and others, subject to scalar constraint $H(x)=0$ and vector constraint $H_{a}(x)=0$. The constraints generate diffeomorphisms in the time and spatial directions respectively, and satisfy the hypersurface deformation algebra as in (\ref{constraint algebra}). We assume that $\{T^{I}(x),P_{J}(x')\}=\delta^{I}_{J}\delta^{(3)}(x,x')$, with $T^{I}, I=0,1,2,3$, being a set of spacetime scalar fields. This implies that $H_{a}(x)$ and $H(x)$ take the forms \cite{Dittrich2,Kuchar:1976yw,Kuchar:1976yx}:
\begin{eqnarray}
	H_{a}(x)&=&P_{I}(x)\partial_{a}T^{I}(x)+H^{R}_{a}(x),\\
	H(x)&=&\mathfrak{H}(P_{I}(x),S_{\mathfrak{n}}(x)),
\end{eqnarray}
where $\mathfrak{n}=1,\cdots,\mathfrak{N}\in\mathbb{N}$, $H^{R}_{a}(x)$ and $S_{\mathfrak{n}}(x)$ are independent of $P_{I}(x)$ and their derivatives. The deparametrization aims to construct a phase space corresponding to the 3+1 decomposition of the spacetime, with $T^{0}$ as the time coordinate and $T^{i}, i=1,2,3$ as the spatial coordinates. In this phase space, the scalar and vector constraints are naturally resolved, and there exists a physical Hamiltonian that generates evolution with respect to $T^{0}$.

To this end, one first needs to deparametrize the constraint subalgebra of $H_{a}(x)$ using $T^{i}(x)$. We assume that the matrix $\{T^{i}(x),H_{a}(x')\}=\partial_{a}T^{i}(x)\delta^{(3)}(x,x')=:J^{i}_{a}(x)\delta^{(3)}(x,x')$ is invertible, so that the phase space functions $T^{i}(x)$ can serve as coordinates for each gauge orbit of the constraints $H_{a}(x)$, and the spacetime scalar fields $T^{i}$ can serve as coordinates for the spatial hypersurface $\Sigma$. Then, for a dynamical field variable $\psi(x)$, there are two methods to construct observables related to it that are invariant under spatial diffeomorphism transformations. The first construction is based on the fact that $\chi^{(T^{j})}:\Sigma\ni (x^{a})\mapsto(\mathfrak{x}^{i}\equiv\!T^{i}(x))\in\prod\limits_{k=1}^{3}Ran(T^{k})=:\tilde{\Sigma}$ is a diffeomorphism, so the phase space function $\tilde{\psi}(\mathfrak{x})\equiv\chi^{(T^{j})}_{*}\psi(\mathfrak{x})$ with a fixed $\mathfrak{x}$ is spatially diffeomorphism-invariant, where the subscript $*$ denotes the pushforward map \cite{Giesel1,Giesel6}. This holds because of $\chi^{(\mathfrak{u}_{*}T^{j})}_{*}\mathfrak{u}_{*}\psi(\mathfrak{x})=(\chi^{(T^{j})}\circ \mathfrak{u}^{-1})_{*}\mathfrak{u}_{*}\psi(\mathfrak{x})=\chi^{(T^{j})}_{*}\psi(\mathfrak{x})$, where $\mathfrak{u}$ is an arbitrary spatial diffeomorphism. By definition, $\tilde{\psi}(\mathfrak{x})$ acquires its physical meaning as the value of the field $\psi$ at point $\mathfrak{x}$ in the $T^{j}$ spatial coordinate system. The second construction relies on the gauge fixing $\mathcal{G}^{i}(x):=T^{i}(x)-\tau^{i}(x)=0$. For a phase space function $f$, by performing a spatial diffeomorphism-invariant extension of $f|_{H_{a}(x)=\mathcal{G}^{i}(x)=0}$, we obtain the phase space function $D_{[f;T^{i}]}(\tau^{j})$, such that $\{D_{[f;T^{i}]}(\tau^{j}),H_{a}(x)\}|_{H_{a}(x')=0}=0$ \cite{Dapor:2013hca,Dittrich1,Dittrich2}. It turns out that observables of the form $D_{[f;T^{i}]}(\tau^{j})$ have the following useful properties \cite{Dittrich1,Dittrich2,Thiemann,Thiemann3}:

\begin{enumerate}[label=(\roman*)]
	\item \label{D1}
	$D_{[f;T^{i}]}(\tau^{j})$ can be expanded as the following series:
	\begin{align}
			&\left.D_{[f;T^{i}]}(\tau^{j})\right|_{H_{a}(x)=0}\nonumber\\	=&\sum_{r=0}^{\infty}\!\frac{1}{r!}\!\int_{\Sigma}\!d^3x_{1}\!\cdots\!\int_{\Sigma}\!d^3x_{r}\!\{\cdots\{f,(J^{-1})^{a_{1}}_{k_{1}}\!(x_{1})\nonumber\\
			\times&\!H_{a_{1}}\!(x_{1})\},\cdots,(J^{-1})^{a_{r}}_{k_{r}}\!(x_{r})H_{a_{r}}\!(x_{r})\}\nonumber\\
			\times&\!(\tau^{k_{1}}\!(x_{1})\!-\!T^{k_{1}}\!(x_{1}))\!\cdots\!(\tau^{k_{r}}\!(x_{r})\!-\!T^{k_{r}}\!(x_{r}))\!|_{H_{a}(x)=0}.
	\end{align}
	\item \label{D2} 
	If $f$ is composed of phase space functions $g_{\mathfrak{m}}$, where $\mathfrak{m}=1,\cdots,\mathfrak{M}\in\mathbb{N}$, then:
	\begin{align}
		&\left.D_{[f(g_{1},\cdots,g_{\mathfrak{M}});T^{i}]}(\tau^{j})\right|_{H_{a}(x)\!=\!0}\nonumber\\
		=&\left.f(D_{[g_{1};T^{i}]}(\tau^{j}),\!\cdots\!,D_{[g_{\mathfrak{M}};T^{i}]}(\tau^{j}))\right|_{H_{a}(x)\!=\!0}.\label{Haproperty2}
	\end{align}	
	\item \label{D3} 
	The Poisson brackets between these observables satisfy the following relation:
	\begin{align}
		&\left.\{D_{[f;T^{i}]}(\tau^{j}),D_{[g;T^{k}]}(\tau^{k})\}\right|_{H_{a}(x)=0}\nonumber\\
		=&\left.{D}_{[\{f,g\}^{\mathfrak{D}};T^{i}]}(\tau^{j})\right|_{H_{a}(x)=0},\label{Haproperty3}
	\end{align}
	where $\{,\}^{\mathfrak{D}}$ denotes the Dirac bracket with respect to the constraints $H_{a}(x)=0$ and the gauge fixing $\mathcal{G}^{i}(x)=0$, which can be written as:
	\begin{align}
		\{f,g\}^{\mathfrak{D}}&\!:=\!\{f,g\}\nonumber\\
		&-\int_{\Sigma}d^3x\{f,(J^{-1})^{a}_{i}(x)H_{a}(x)\}\{T^{i}(x),g\}\nonumber\\
		&+\int_{\Sigma}d^3x\{f,T^{i}(x)\}\{(J^{-1})^{a}_{i}(x)H_{a}(x),g\}.
	\end{align}
\end{enumerate}

The second construction can be generalized to arbitrary first-class constraint systems. Moreover, since $\tilde{\psi}(\mathfrak{x})|_{H_{a}(x')=\mathcal{G}^{i}(x')=0}=\chi^{(T^{j})}_{*}\psi(\chi^{(T^{j})}(x))|_{H_{a}(x')=\mathcal{G}^{i}(x')=0}$, where $\chi^{(\tau^{j})}(x)=\mathfrak{x}$, one has $\tilde{\psi}(\mathfrak{x})|_{H_{a}(x')=0}=D_{[\chi^{(T^{k})}_{*}\psi(\chi^{(T^{k})}(x));T^{i}]}(\tau^{j})|_{H_{a}(x')=0}$. One can use this relation to conveniently compute the Poisson brackets between observables in the first construction by the corresponding observables in the second construction. For example, if the Poisson bracket between the spatial scalar field $\phi(x)$ and its conjugate momentum $p_{\phi}(x')$ is $\{\phi(x),p_{\phi}(x')\}=\mathsf{c}\delta^{(3)}(x,x')$, then \cite{Dittrich1,Dittrich2}:
\begin{equation}
	\begin{split}
		&\left.\{\tilde{\phi}(\mathfrak{x}),\tilde{p}_{\phi}(\mathfrak{x}')\}\right|_{H_{a}(x'')=0}\\
		=\!&\{\!D_{[\phi\!(x);T^{i}]}\!(\tau^{j}),\!D_{[p_{\phi}\!(x')|\!\det(\!J\!)\!(x')\!|^{-1};T^{i}]}\!(\tau^{j})\}\!|_{H_{a}\!(x'')\!=\!0}\\
		=&\left.D_{[\{\phi(x),p_{\phi}(x')|\det(J)(x')|^{-1}\}^{\mathfrak{D}};T^{i}]}(\tau^{j})\right|_{H_{a}(x'')=0}\\
		=&\mathsf{c}\delta^{(3)}(x,x')\left|\det(\partial_{a}\tau^{k})(x')\right|^{-1}\\
		=&\mathsf{c}\delta^{(3)}(\mathfrak{x},\mathfrak{x}').
	\end{split}
\end{equation}
It is easy to show that the Poisson brackets of the spatial diffeomorphism-invariant observables corresponding to the basic variables other than $(T^{i}(x),P_{j}(x'))$ have a similar form. By setting $\tilde{T}^{i}(\mathfrak{x})|_{H_{a}(x')=0}=\mathfrak{x}^{i}$ and $\tilde{P}_{j}(\mathfrak{x}'')|_{H_{a}(x')=0}=[-\tilde{P}_{0}(\mathfrak{x}')\partial_{j}\tilde{T}^{0}(\mathfrak{x}')-\tilde{H}^{R}_{j}(\mathfrak{x}')]|_{H_{a}(x'')=0}$, we get a set of canonical coordinates $(\tilde{T}^{0}(\mathfrak{x}),\tilde{P}_{0}(\mathfrak{x}')), (\tilde{q}_{ij}(\mathfrak{x}),\tilde{p}^{kl}(\mathfrak{x}')),\cdots$ on the reduced phase space $\Gamma_{Diff}$ solving the vector constraints \cite{Giesel1,Giesel6}. Furthermore, since the constraints $H_{a}(x)=H(x)=0,\forall x\in\Sigma$, is equivalent to the constraints $H_{a}(x)=\tilde{H}(\mathfrak{x})=0,\forall x\in\Sigma$ and $\forall\mathfrak{x}\in\tilde{\Sigma}$, the spatial diffeomorphism-invariant scalar constraint $\tilde{H}(\mathfrak{x})$ and the vector constraint $H_{a}(x)$ form a first-class constraint system in $\Gamma_{pre}$. This implies that the constraints $\tilde{H}(\mathfrak{x})$ are of first-class in $\Gamma_{Diff}$ \cite{Dittrich1,Dittrich2}.

Next, one needs to deparametrize the constraints $\tilde{H}(\mathfrak{x})=\mathfrak{H}(\tilde{P}_{0}(\mathfrak{x}),-\tilde{P}_{0}(\mathfrak{x})\partial_{i}\tilde{T}^{0}(\mathfrak{x})-\tilde{H}^{R}_{i}(\mathfrak{x}),\tilde{S}_{\mathfrak{n}}(\mathfrak{x}))=:\mathfrak{H}'(\tilde{P}_{0}(\mathfrak{x}),\tilde{L}_{\mathfrak{r}}(\mathfrak{x}))=0$ in $\Gamma_{Diff}$ using $\tilde{T}^{0}(\mathfrak{x})$, where $\tilde{L}_{\mathfrak{r}}\in\{\partial_{i}\tilde{T}^{0},\tilde{H}^{R}_{i},\tilde{S}_{\mathfrak{n}}\}$. We assume that the matrix $\{\tilde{T}^{0}(\mathfrak{x}),\tilde{H}(\mathfrak{x'})\}=(\partial\mathfrak{H}'/\partial\tilde{P}_{0})(\mathfrak{x})\delta^{(3)}(\mathfrak{x},\mathfrak{x}')$ is invertible, so that the phase space function $\tilde{T}^{0}(\mathfrak{x})$ can serve as the coordinate for each gauge orbit of the constraints $\tilde{H}(\mathfrak{x})$, and the spacetime scalar field $\tilde{T}^{0}$ can serve as the time coordinate. To identify $\tilde{T}^{0}$ as the time in the 3+1 decomposition of spacetime, we gauge fix $\tilde{\mathcal{G}}^{0}(\mathfrak{x}):=\tilde{T}^{0}(\mathfrak{x})-\tau=0$, where $\tau$ is a constant. For the given phase space function $\tilde{\psi}(\mathfrak{x})$, a gauge-invariant extension of $\tilde{\psi}(\mathfrak{x})|_{\tilde{H}(\mathfrak{x'})=\tilde{\mathcal{G}}^{0}(\mathfrak{x'})=0}$ gives the Dirac observable $O_{[\tilde{\psi}(\mathfrak{x});\tilde{T}^{0}]}(\tau)=:\bar{\psi}(\mathfrak{x},\tau)$, satisfying $\{\bar{\psi}(\mathfrak{x},\tau),\tilde{H}(\mathfrak{x'})\}|_{\tilde{H}(\mathfrak{x''})=0}=0$. The symbol $O_{[\cdot;\tilde{T}^{0}]}(\tau)$ indicates that it possesses properties completely analogous to those of \ref{D1}, \ref{D2}, and \ref{D3} \cite{Dittrich1,Dittrich2,Thiemann,Thiemann3}. Therefore, the Poisson bracket between $\bar{\phi}(\mathfrak{x},\tau)$ and $\bar{p}_{\phi}(\mathfrak{x'},\tau)$ can be calculated as \cite{Dittrich1,Dittrich2}:
\begin{align}
	&\left.\{\bar{\phi}(\mathfrak{x},\tau),\bar{p}_{\phi}(\mathfrak{x'},\tau)\}\right|_{\tilde{H}(\mathfrak{x}'')=0}\nonumber\\
	=&\left.{O}_{[\{\tilde{\phi}(\mathfrak{x}),\tilde{p}_{\phi}(\mathfrak{x}')\}^{\mathscr{D}};\tilde{T}^{0}]}(\tau)\right|_{\tilde{H}(\mathfrak{x}'')=0}\nonumber\\
	=&\mathsf{c}\delta^{(3)}(\mathfrak{x},\mathfrak{x}'),
\end{align}
where $\{,\}^{\mathscr{D}}$ denotes the Dirac bracket with respect to the constraints $\tilde{H}(\mathfrak{x})=0$ and the gauge fixing $\tilde{\mathcal{G}}^{0}(\mathfrak{x})=0$, which can be expressed as:
\begin{align}
	\{\tilde{f},\tilde{g}\}^{\mathscr{D}}&:=\{\tilde{f},\tilde{g}\}\nonumber\\
	&-\int_{\tilde{\Sigma}}d^3\mathfrak{x}\{\tilde{f},(\frac{\partial\mathfrak{H}'}{\partial\tilde{P}_{0}})^{-1}(\mathfrak{x})\tilde{H}(\mathfrak{x})\}\{\tilde{T}^{0}(\mathfrak{x}),\tilde{g}\}
	\nonumber\\
	&+\int_{\tilde{\Sigma}}d^3\mathfrak{x}\{\tilde{f},\tilde{T}^{0}(\mathfrak{x})\}\{(\frac{\partial\mathfrak{H}'}{\partial\tilde{P}_{0}})^{-1}(\mathfrak{x})\tilde{H}(\mathfrak{x}),\tilde{g}\}.
\end{align} 
It is obvious that the Poisson brackets between the Dirac observables corresponding to the basic variables, apart from $(\tilde{T}^{0}(\mathfrak{x}),\tilde{P}_{0}(\mathfrak{x}'))$, remain invariant. Moreover, the condition $\partial\mathfrak{H}'(\tilde{P}_{0},\tilde{L}_{\mathfrak{r}})/\partial\tilde{P}_{0}\ne 0$ implies $\partial\mathfrak{H}'(\bar{P}_{0},\bar{L}_{\mathfrak{r}})/\partial\bar{P}_{0}\ne 0$. Hence, the equation $\mathfrak{H}'(\bar{P}_{0},\bar{L}_{\mathfrak{r}})=0$ can be locally solved as $\bar{P}_{0}=-\bar{\mathfrak{h}}(\bar{L}_{\mathfrak{r}})$, according to the implicit function theorem. Thus, by setting $\bar{T}^{0}(\mathfrak{x},\tau)|_{\tilde{H}(\mathfrak{x}')=0}=\tau$, we obtain a set of canonical coordinates $(\bar{q}_{ij}(\mathfrak{x},\tau),\bar{p}^{kl}(\mathfrak{x}',\tau)),\cdots$ on the reduced phase space $\Gamma_{phy}$ solving all constraints \cite{Giesel1,Giesel6}.

The evolution of the Dirac observable $\bar{\psi}(\mathfrak{x},\tau)$ on phase space $\Gamma_{phy}$ with respect to $\tau$ can be calculated as \cite{Giesel2}:
\begin{align}
	&\left.\partial_{\tau}\bar{\psi}(\mathfrak{x},\!\tau)\right|_{\Gamma_{phy}}\nonumber\\
	=\!&\sum\limits_{r=1}^{\infty}\!\frac{1}{(r\!-\!1)!}\!\int_{\tilde{\Sigma}}\!d^3\!\mathfrak{x}_{1}\!\cdots\!\int_{\tilde{\Sigma}}\!d^3\!\mathfrak{x}_{r}\!\{\cdots\{\tilde{\psi}\!(\mathfrak{x})\!,\!(\!\frac{\partial\mathfrak{H}'}{\partial\tilde{P}_{0}}\!)^{-\!1}\!(\!\mathfrak{x}_{1}\!)\nonumber\\
	\times&\tilde{H}(\mathfrak{x}_{1})\},\cdots,(\!\frac{\partial\mathfrak{H}'}{\partial\tilde{P}}\!)^{-\!1}(\mathfrak{x}_{r})\tilde{H}(\mathfrak{x}_{r})\}\nonumber\\
	\times&\left.(\tau-\tilde{T}^{0}(\mathfrak{x}_{1}))\cdots(\tau-\tilde{T}^{0}(\mathfrak{x}_{r-1}))\right|_{\tilde{H}(\mathfrak{x}')=0}\nonumber\\
	=&\left.O_{[\{\tilde{\psi}(\mathfrak{x}),\int_{\tilde{\Sigma}}d^3\mathfrak{x}'(\frac{\partial\mathfrak{H}'}{\partial\tilde{P}_{0}})^{-1}(\mathfrak{x}')\tilde{H}(\mathfrak{x}')\};\tilde{T}^{0}]}(\tau)\right|_{\tilde{H}(\mathfrak{x}'')=0}\nonumber\\
	=&\!\left.O_{\![\!\sum\limits_{\mathfrak{r}}\!\int_{\tilde{\Sigma}}\!d^3\!\mathfrak{x}'\!(\!\frac{\partial\mathfrak{H}'}{\partial\tilde{P}_{0}}\!)^{-\!1}(\!\mathfrak{x}'\!)\frac{\partial\mathfrak{H}'}{\partial\tilde{L}_{\mathfrak{r}}}(\!\mathfrak{x}'\!)\{\!\tilde{\psi}\!(\!\mathfrak{x}\!),\tilde{L}_{\mathfrak{r}}\!(\!\mathfrak{x}'\!)\!\};\tilde{T}^{0}]}\!(\tau)\right|_{\tilde{H}(\mathfrak{x}'')\!=\!0}\nonumber\\
	=&\!\sum_{\mathfrak{r}}\!\int_{\tilde{\Sigma}}\!d^3\mathfrak{x}'(\!\!\frac{\partial\mathfrak{H}'(\bar{P}_{0},\bar{L}_{\mathfrak{r}})}{\partial\bar{P}_{0}}\!)^{-\!1}\!(\mathfrak{x}',\!\tau)\!\frac{\partial\mathfrak{H}'(\bar{P}_{0},\bar{L}_{\mathfrak{r}})}{\partial\bar{L}_{\mathfrak{r}}}(\mathfrak{x}',\!\tau)\nonumber\\
	\times&\left.\{\bar{\psi}(\mathfrak{x},\tau),\bar{L}_{\mathfrak{r}}(\mathfrak{x}',\tau)\}\right|_{\Gamma_{phy}}\nonumber\\
	=&\!\left.\sum_{\mathfrak{r}}\!\int_{\tilde{\Sigma}}\!d^3\!\mathfrak{x}'\!\frac{\partial\bar{\mathfrak{h}}}{\partial\bar{L}_{\mathfrak{r}}}\!(\mathfrak{x}',\!\tau)\!\{\!\bar{\psi}\!(\mathfrak{x},\!\tau),\!\bar{L}_{\mathfrak{r}}\!(\mathfrak{x}',\!\tau)\!\}\right|_{[\!\bar{P}_{0}\!+\!\bar{\mathfrak{h}}]\!(\mathfrak{x}'',\!\tau)\!=\!0}\nonumber\\
	=&\left.\{\bar{\psi}(\mathfrak{x},\tau),\int_{\tilde{\Sigma}}d^3\mathfrak{x}'\bar{\mathfrak{h}}(\mathfrak{x}',\tau)\}\right|_{[\bar{P}_{0}+\bar{\mathfrak{h}}](\mathfrak{x}'',\tau)=0},
\end{align}
where, in the first step, the Dirac observable $\bar{\psi}(\mathfrak{x},\tau)$ is expanded into an expression analogous to \ref{D1} and differentiated with respect to $\tau$; subsequently, the commutativity of the Hamiltonian vector field of $(\partial\mathfrak{H}'/\partial\tilde{P}_{0})^{-1}(\mathfrak{x})\tilde{H}(\mathfrak{x})$ on the constraint surface $\tilde{H}(\mathfrak{x})=0$ is employed to reorder the terms. In steps two to four, properties \ref{D1}, \ref{D2} and \ref{D3} of the Dirac observable are used. In the penultimate step, we restricted the phase space $\Gamma_{phy}$ to the region defined by $[\bar{P}_{0}+\bar{\mathfrak{h}}](\mathfrak{x},\tau)=0$. Therefore, in this region, the physical Hamiltonian $H_{phy}\equiv\int_{\tilde{\Sigma}}d^3\mathfrak{x}\bar{\mathfrak{h}}(\mathfrak{x},\tau)$ generates the evolution of $\bar{\psi}(\mathfrak{x},\tau)$ relative to $\tau$.

\subsection{Relation to the Gauge-Fixed Scheme}
The assumption that the Poisson brackets between $T^{I}(x)$ and the constraints are invertible implies that not only $T^{I}(x)$ can serve as coordinates on the gauge orbits within the phase space, but also $T^{I}$ can be interpreted as spacetime coordinates. In the deparametrization scheme, we mainly exploit the former property. Here, one constructs Dirac observables by performing a gauge-invariant extension of the dynamical variables on the gauge-fixed section defined by $T^{I}(x)=\tau^{I}(x)$, and then varies the section to yield the notion of relational evolution of the Dirac observables. In the gauge-fixed scheme, one primarily makes use of the latter property. Here, by substituting $T^{I}$ as spacetime coordinates into the equations of motion, the evolution of the dynamical variables with respect to $T^{I}$ is obtained. We are going to derive the physical Hamiltonian for relative evolution by gauge-fixing the dynamical variables to spacetime coordinates, and compare it with the deparametrization scheme. To this end, we gauge fix $\mathcal{G}^{0}(x)\equiv\!T^{0}(x)-\tau(t)=0$ and $\mathcal{G}^{i}(x)\equiv\!T^{i}(x)-\tau^{i}(x)=0$, where $d\tau/d{t}\ne0$ and the matrix $M^{i}_{a}(x)\equiv\partial_{a}\tau^{i}(x)$ is invertible \cite{Giesel1,Giesel6}. To ensure that the gauge-fixings are maintained during the evolution, we require:
\begin{align}
	0=&\left.\frac{d\mathcal{G}^{0}}{dt}\right|_{\Gamma_{\tau}}=\left.(\{T^{0},H_{can}\}-\frac{d\tau}{dt})\right|_{\Gamma_{\tau}}\nonumber\\
	=&\left.N\dfrac{\partial{\mathfrak{H}}}{\partial{P_{0}}}\right|_{\Gamma_{\tau}}-\frac{d\tau}{dt},\\
	0=&\left.\frac{d\mathcal{G}^{i}}{dt}\right|_{\Gamma_{\tau}}=\left.\{T^{i},H_{can}\}\right|_{\Gamma_{\tau}}\nonumber\\
	=&\left.(N\dfrac{\partial{\mathfrak{H}}}{\partial{P_{i}}}+M^{i}_{a}N^{a})\right|_{\Gamma_{\tau}},
\end{align}
where $H_{can}\equiv\int_{\Sigma}d^{3}xNH+N^{a}H_{a}$, the gauge-fixed section $\Gamma_{\tau}$ is defined by $H=H_{a}=\mathcal{G}^{I}=0$, and the index $x$ is omitted. Thus, one get $\left.\partial{\mathfrak{H}}/\partial{P}_{0}\right|_{\Gamma_{\tau}}\ne0$, $N=\left.(d{\tau}/d{t})(\partial{\mathfrak{H}}/\partial{P}_{0})^{-1}\right|_{\Gamma_{\tau}}$, and $N^{a}=-\left.N(M^{-1})^{a}_{k}(\partial{\mathfrak{H}}/\partial{P_{k}})\right|_{\Gamma_{\tau}}$. Moreover, from the gauge-fixed conditions, we obtain
\begin{align}
	0&=(P_{k}+(M^{-1})^{a}_{k}H^{R}_{a}\left.)\right|_{\Gamma_{\tau}},\\
	0&=\left.\mathfrak{H}(P_{0},-(M^{-1})^{a}_{k}H^{R}_{a},S_{\mathfrak{n}})\right|_{\Gamma_{\tau}}.\label{HC}
\end{align}
With the help of the implicit function theorem, we can locally solve $P_{0}=-\mathfrak{h}(-(M^{-1})^{a}_{k}H^{R}_{a},S_{\mathfrak{n}})$ from (\ref{HC}). 

The above results imply that we only need to consider the evolution of the dynamical variable $\psi$, which is independent of $T^{I}$ and $P_{J}$. A direct calculation gives
\begin{align}
	{}&\frac{d\psi}{d\tau}=(\frac{d\tau}{dt})^{-1}\frac{d\psi}{dt}=(\frac{d\tau}{dt})^{-1}\{\psi,H_{can}\}\nonumber\\
	\simeq&(\frac{d\tau}{dt})^{-1}\!\int_{\Sigma}d^{3}x[N(x)\!\sum_{\mathfrak{n}}\frac{\partial{\mathfrak{H}}}{\partial{S_{\mathfrak{n}}}}(x)\{\psi,S_{\mathfrak{n}}(x)\}\nonumber\\
	+&\!N^{a}(x)\{\psi,H^{R}_{a}(x)\}]\nonumber\\
	\simeq&\int_{\Sigma}d^{3}x[(\dfrac{\partial{\mathfrak{H}}}{\partial{P}_{0}})^{-1}\sum_{\mathfrak{n}}\frac{\partial{\mathfrak{H}}}{\partial{S_{\mathfrak{n}}}}(x)\{\psi,S_{\mathfrak{n}}(x)\}\nonumber\\
	-&(M^{-1})^{a}_{i}(x)(\dfrac{\partial{\mathfrak{H}}}{\partial{P}_{0}})^{-1}(x)\dfrac{\partial{\mathfrak{H}}}{\partial{P_{i}}}(x)\{\psi,H^{R}_{a}(x)\}]\nonumber\\
	\simeq\!&\int_{\Sigma}\!d^{3}\!x[\sum_{\mathfrak{n}}\!\frac{\partial{\mathfrak{h}}}{\partial{S_{\mathfrak{n}}}}\!(x)\{\!\psi,\!S_{\mathfrak{n}}\!(x)\!\}\!+\!\dfrac{\partial{\mathfrak{h}}}{\partial{(-\!(\!M^{-\!1}\!)^{a}_{i}\!H^{R}_{a})}}\!(x)\nonumber\\
	\times&\{\psi,(-(M^{-1})^{a}_{i}(x)C^{R}_{a}(x))\}]\Big|_{P_{0}+\mathfrak{h}=0}\nonumber\\
	\simeq&\left.\{\psi,\int_{\Sigma}d^{3}x\mathfrak{h}(x)\}\right|_{P_{0}+\mathfrak{h}=0},
\end{align}
where the symbol $\simeq$ means that the equation holds when $H=H_{a}=\mathcal{G}^{J}=d{\mathcal{G}}^{J}/dt=0$. 
In the penultimate step, we restrict the section $\Gamma_{\tau}$ to the region where $P_{0}+\mathfrak{h}=0$. In addition, based on the construction of Dirac observables in the previous section, we have $\left.\int_{\tilde{\Sigma}}d^3\mathfrak{x}\bar{P}_{0}(\mathfrak{x},\tau)\right|_{\Gamma_{phy}}=\int_{\tilde{\Sigma}}d^3\mathfrak{x}P_{0}((\chi^{(\tau^{j})})^{-\!1}(\mathfrak{x}))|\det(J)((\chi^{(\tau^{j})})^{-1}(\mathfrak{x}))|^{-\!1}\big|_{\Gamma_{\tau}}\\=\left.\int_{\Sigma}d^3xP_{0}(x)\right|_{\Gamma_{\tau}}$. Therefore, as long as the regions, where $\bar{P}_{0}$ and $P_{0}$ are solved respectively, are consistent under natural projection, the dynamics of deparametrization scheme is equivalent to the dynamics of the gauge-fixing scheme, i.e., $\left.\int_{\tilde{\Sigma}}d^3\mathfrak{x}\bar{\mathfrak{h}}(\mathfrak{x},\tau)\right|_{\Gamma_{phy}}=\left.\int_{\Sigma}d^{3}x\mathfrak{h}(x)\right|_{\Gamma_{\tau}}$.

\subsection{Spherically Symmetric Reduction}\label{Reduction}

We consider the spherically symmetric model of vector-tensor theory to carry out the deparametrization. In the model the space takes the form $\Sigma=\mathcal{I}\times{\mathbb{S}^{2}}$, where $\mathcal{I}$ is a 1-dimensional manifold, $\mathbb{S}^{2}$ is the orbital sphere of the spherical symmetry group $\mathrm{SO}(3)$, and we define the spherical coordinates $(x^{a})=(x,\theta,\phi)$. By requiring that $g_{\mu\nu}$ and $Y_{\mu}$ remain invariant under any spherically symmetric transformation, we obtain their spherically symmetric forms respectively as:
\begin{equation}\label{symmetry1}
	\begin{split}
		g_{\mu\nu}dx^{\mu}dx^{\nu}\!=\!&-N^2dt^2\!+q_{xx}(dx\!+\!N^{x}dt)^2\\
		+&q_{\theta\theta}(d\theta^2\!+\!\sin^2\theta{d}\phi^2),\\
		Y_{\mu}dx^{\mu}\!=\!&\sqrt{(q_{xx})^{-1}Y_{x}Y_{x}\!-\!\check{Y}}Ndt\\
		+&Y_{x}(dx\!+\!N^{x}dt),
	\end{split}
\end{equation}
where $q_{xx}$, $q_{\theta\theta}$, $N$, $N^{x}$, $Y_{x}$, and $\check{Y}$ depend only on $x$ and $t$. Combining (\ref{momentum}) and (\ref{symmetry1}), we get:
\begin{equation}
	\pi^{Y}=:\frac{\sin\theta}{4\pi}p^{Y}(x),\qquad\pi^{a}\partial_{a}=:\frac{\sin\theta}{4\pi}p^{x}(x)\partial_{x}.
\end{equation}
In addition, requiring $A^{i}_{a}$ and $E^{a}_{i}$ to remain invariant under spherically symmetric transformations, up to a suitable gauge transformation \cite{Bojowald2,Harnad:1979in,Brodbeck:1996ma}, their spherically symmetric forms respectively read:
\begin{equation}\label{sphericallyAE}
	\begin{split}
		{}&A:=A^i_a\tau_idx^a\\
		=&A_x(x)\tau_3dx+(A_1(x)\tau_1+A_2(x)\tau_2)d\theta\\
		+&(A_1(x)\tau_2-A_2(x)\tau_1)\sin\theta{d\phi}+\cos\theta\tau_3{d\phi},\\
		{}&E:=E^a_i\tau^i\partial_a\\
		=&E^x(x)\tau_3\sin\theta\partial_{x}+(E^1(x)\tau_1+E^2(x)\tau_2)\sin\theta\partial_{\theta}\\
		+&(E^1(x)\tau_2-E^2(x)\tau_1)\partial_{\phi}.
	\end{split}
\end{equation}

The phase space $\Gamma_{sph}$ of the spherically symmetric model is a subset of the full theory's phase space, where all field components encoding the dynamical information depend only on $x$. Thus the vector-tensor gravity theory is reduced to a 1+1-dimensional field theory under spherical symmetry. Under radial coordinate transformations, $E^{x}$, $p^{x}$, $\check{Y}$, $A_{1}$ and $A_{2}$ transform as scalar fields, while $A_{x}$, $Y_{x}$, $\pi^{Y}$, $E^{1}$ and $E^{2}$ transform as scalar density fields with weight 1. By restricting the original symplectic form to the spherically symmetric dynamic variables, we obtain the symplectic form on $\Gamma_{sph}$ as:
\begin{align}
		&\int_{\Sigma}\!d^3x\!\bigg(\!\frac{1}{8G\pi\gamma}\delta\!E^a_i\!\wedge\!\delta \!A^i_a\!+\!\delta\!\pi^{Y}\!\wedge\!\delta{\check{Y}}\!+\!\delta\pi^a\!\wedge\!\delta\!{Y_a}\!\bigg) \nonumber\\
		=\!&\int_{\mathcal{I}}\!dx\!\bigg(\!\frac{1}{2G\gamma}\delta\!E^x\!\wedge\!\delta \!A_x\!+\!\frac{1}{G\gamma}(\delta\!E^1\!\wedge\!\delta\!A_1\!+\!\delta\!E^2\!\wedge\!\delta \!A_2)\nonumber\\
		+\!&\delta\!{p^{Y}}\!\wedge\!\delta\!{\check{Y}}+\delta\!{p^{x}}\!\wedge\!\delta\!{Y_{x}}\!\bigg).
\end{align}
Thus, the nontrivial Poisson brackets between the basic variables are:
\begin{equation}
\begin{split}
		\{A_x(x),E^x(x')\}&=2G\gamma\delta(x,x'),\\ \{A_s(x),E^t(x')\}&=G\gamma\delta^{t}_{s}\delta(x,x'),\\
		\{\check{Y}(x),p^{Y}(x')\}&=\delta(x,x'),\\ 
		\{Y_{x}(x),p^{x}(x')\}&=\delta(x,x'),
\end{split}
\end{equation}
where $s,t\in\{1,2\}$.

By restricting the Gauss constraint to spherically symmetric dynamic variables, we get the reduced Gauss constraint on $\Gamma_{sph}$ as \cite{Bojowald1,Chiou}:
\begin{equation}\label{sphericallyGaussian}
	G[\lambda]\!=\!\int_{\mathcal{I}}\!dx\!\frac{\lambda}{2G\gamma}\!(\partial_{x}E^x\!+\!2A_1E^2\!-\!2A_2E^1),
\end{equation}
where $\lambda$ is a smearing function. It generates the following gauge transformations:
\begin{equation}
	\begin{split}
		\exp(\{\cdot,G[\lambda]\})\cdot{A}=&e^{\lambda\tau_{3}}Ae^{-\lambda\tau_{3}}+e^{\lambda\tau_{3}}de^{-\lambda\tau_{3}},\\
		\exp(\{\cdot,G[\lambda]\})\cdot{E}=&e^{\lambda\tau_{3}}Ee^{-\lambda\tau_{3}}.
	\end{split}
\end{equation}
Thus, the $\mathrm{SU}(2)$ gauge transformations are reduced to the $\mathrm{U}(1)$ gauge transformations. One can gauge-fix $E^2(x)=0$ and solve $A_{2}(x)=\partial_{x}E^{x}/(2E^1)(x)$ from (\ref{sphericallyGaussian}) to obtain the phase space $\Gamma_{sph}^{G}$ without the Gauss constraint \cite{Zhang}. However, there is still a residual gauge freedom \cite{Chiou,Ashtekar6}, given by the transformation $\Pi:(A_1(x),E^1(x))\to(-A_1(x),-E^1(x))$. To simplify the quantum theory, we will first solve the local constraints and then require the physical state to be invariant under the action of the operator $\hat{\Pi}$ corresponding to $\Pi$.

By spherically symmetric reduction, the reduced spin connection and $\tilde{K}^{i}_{a}$ can be calculated as:
\begin{equation}
	\begin{split}
		&\Gamma_{spin}:=\Gamma^{i}_{a}\tau_{i}dx^{a}\\
		=&\!\frac{\partial_{x}E^x}{2E^1}(x)\tau_2d\theta\!-\!\frac{\partial_{x}E^x}{2E^1}(x)\sin\theta\tau_1d\phi\!+\!\cos\theta\tau_3d\phi,\\
		&\gamma\tilde{K}:=\gamma\tilde{K}^{i}_{a}\tau_{i}dx^{a}=A-\Gamma_{spin}\\
		=&A_x(x)\tau_3dx+A_1(x)\tau_1d\theta+A_1(x)\tau_2\sin\theta{d\phi}.
	\end{split}
\end{equation}
For convenience \cite{Bojowald3}, we define $\tilde{K}_{x}:=A_{x}/{\gamma}$, $\tilde{K}_{\varphi}:=A_{1}/\gamma$ and $E^{\varphi}:=E^1$. Then the nontrivial Poisson brackets on $\Gamma_{sph}^{G}$ read
\begin{equation}
	\begin{split}
		\{\tilde{K}_{x}(x),E^x(x')\}&=2G\delta(x,x'),\\ \{\tilde{K}_{\varphi}(x),E^{\varphi}(x')\}&=G\delta(x,x'),\\
		\{\check{Y}(x),p^{Y}(x')\}&=\delta(x,x'),\\ 
		\{Y_{x}(x),p^{x}(x')\}&=\delta(x,x').
	\end{split}
\end{equation}

To consider the spherically symmetric reductions of the scalar and vector constraints, we decompose the curvature $F=dA+(1/2)A\wedge{A}$ as:
\begin{align}\label{Fdecompose}
	F=R_{spin}+\tilde{F}+\Gamma_{spin}\wedge\gamma\tilde{K},
\end{align}
where
\begin{align}
	{}&R_{spin}:=d\Gamma_{spin}+\frac{1}{2}\Gamma_{spin}\wedge\Gamma_{spin}\nonumber\\
	=\!&\partial_{x}(\frac{\partial_{x}E^x}{2E^{\varphi}})\tau_2dx\!\wedge\!d\theta\!-\!\partial_{x}(\frac{\partial_{x}E^x}{2E^{\varphi}})\sin\theta\tau_1dx\!\wedge\!d\phi\nonumber\\
	+&((\frac{\partial_{x}E^x}{2E^{\varphi}})^2-1)\sin\theta\tau_3d\theta\wedge{d}\phi,
\end{align}
and
\begin{align}
	&\tilde{F}:=d(\gamma\tilde{K})+\frac{1}{2}(\gamma\tilde{K})\wedge(\gamma\tilde{K})\nonumber\\
	=&\gamma\partial_{x}\tilde{K}_{\varphi}\tau_1dx\!\wedge\!{d}\theta\!+\!\gamma\partial_{x}\tilde{K}_{\varphi}\sin\theta\tau_2dx\!\wedge\!{d}\phi\nonumber\\
	+&\!\gamma\tilde{K}_{\varphi}\cos\theta\tau_2d\theta\!\wedge\!{d}\phi+\gamma^2\tilde{K}_{x}\tilde{K}_{\varphi}\tau_2dx\wedge{d}\theta\nonumber\\
	-&\!\gamma^2\!\tilde{K}_{x}\!\tilde{K}_{\varphi}\!\sin\theta\tau_1dx\!\wedge\!d\phi\!+\!\gamma^2\!(\tilde{K}_{\varphi})^2\!\sin\theta\tau_3d\theta\!\wedge\!d\phi,
\end{align}
with
\begin{align}
	&\Gamma_{spin}\wedge\gamma\tilde{K}\nonumber\\
	=&\!-\!\gamma\!\frac{\partial_{x}E^x}{2E^{\varphi}}\!\tilde{K}_{x}\tau_1dx\!\wedge\!d\theta\!-\!\gamma\!\frac{\partial_{x}E^x}{2E^{\varphi}}\!\tilde{K}_{x}\sin\theta{\tau_{2}}dx\!\wedge\!d\phi\nonumber\\
	-&\gamma\tilde{K}_{\varphi}\cos\theta\tau_2d\theta\wedge{d}\phi.
\end{align}
By restricting to the spherically symmetric dynamic variables, we obtain the reduced vector constraints as:
\begin{align}
	\mathcal{C}_{x}[N^{x}]\!=&\!\int_{\mathcal{I}}dxN^{x}(\frac{1}{2G}(2E^{\varphi}\partial_{x}\tilde{K}_{\varphi}\!-\!\tilde{K}_{x}\partial_{x}E^{x})\nonumber\\
	+&p^{Y}\partial_{x}\check{Y}\!-\!Y_{x}\partial_{x}p^{x}),\label{sCx}
\end{align}
where $2E^{\varphi}\partial_{x}\tilde{K}_{\varphi}=(\tilde{F}_{x\theta}^{1}E^{\theta}_{1}+\tilde{F}_{x\phi}^{2}E^{\phi}_{2})|_{\theta=\pi/2}$ and $-\tilde{K}_{x}\partial_{x}E^{x}=((\Gamma_{spin}\wedge\tilde{K})_{x\theta}^{1}E^{\theta}_{1}+(\Gamma_{spin}\wedge\tilde{K})_{x\phi}^{2}E^{\phi}_{2})|_{\theta=\pi/2}$. It turns out that $\mathcal{C}_{x}[N^{x}]$ generates radial spatial diffeomorphism transformations. Similarly, all the terms in (\ref{nine terms}) can be calculated by the spherically symmetric reduction. The first term is reduced to
\begin{align}
	\mathcal{C}_{1}[N]\!=&\!\int_{\mathcal{I}}\!dx\!N\frac{1\!+\!\omega{\check{Y}}}{2G}|E^{\varphi}|^{-\!1}|E^{x}|^{-\!\frac{1}{2}}\nonumber\\
	\times&\!(\gamma^2(\tilde{K}_{\varphi})^2(E^{\varphi})^{2}+2\gamma^2\!\tilde{K}_{x}\tilde{K}_{\varphi}E^{x}E^{\varphi}\nonumber\\
	+&\!(E^{\varphi})^{2}((\!\frac{\partial_{x}E^x}{2E^{\varphi}}\!)^{2}\!-\!1)\!+\!2E^{x}E^{\varphi}\partial_{x}(\!\frac{\partial_{x}E^x}{2E^{\varphi}}\!)),\label{sC1}
\end{align}
where $\gamma^2(\tilde{K}_{\varphi})^2(E^{\varphi})^{2}=(\tilde{F}^{3}_{\theta\phi}E^{\theta}_{1}E^{\phi}_{2})|_{\theta=\pi/2}$, $2\gamma^2\!\tilde{K}_{x}\!\tilde{K}_{\varphi}\!E^{x}\!E^{\varphi}\!=\!(\tilde{F}^{1}_{\phi{x}}\!E^{\phi}_{2}\!E^{x}_{3}\!+\!\tilde{F}^{2}_{{x}\theta}\!E^{x}_{3}\!E^{\theta}_{1})\!|_{\theta=\pi/2}$, and the second row of (\ref{sC1}) comes from $R_{spin}$. The second term is reduced to
\begin{align}
\mathcal{C}_{2}[N]\!=\!&\int_{\mathcal{I}}\!dx\!\frac{-N}{2G}(\dfrac{1}{1\!+\!\omega{\check{Y}}}\!+\!\gamma^2(1\!+\!\omega{\check{Y}}))|E^{\varphi}|^{-\!1}\nonumber\\
\times&|E^{x}|^{-\!\frac{1}{2}}((\tilde{K}_{\varphi})^2(E^{\varphi})^{2}\!+\!2\tilde{K}_{x}\tilde{K}_{\varphi}E^{x}E^{\varphi}),\label{sC2}
\end{align}
where $(\tilde{K}_{\varphi})^2(E^{\varphi})^{2}=(\tilde{K}^{1}_{\theta}\tilde{K}^{2}_{\phi}E^{\theta}_{1}E^{\phi}_{2})|_{\theta=\pi/2}$, $2\tilde{K}_{x}\!\tilde{K}_{\varphi}\!E^{x}\!E^{\varphi}\!=\!(\tilde{K}^{2}_{\phi}\!\tilde{K}^{3}_{x}\!E_{2}^{\phi}\!E_{3}^{x}\!+\!\tilde{K}^{1}_{\theta}\!\tilde{K}^{3}_{x}\!E_{1}^{\theta}\!E_{3}^{x})\!|_{\theta=\pi/2}$. The third and fourth term are respectively reduced to
\begin{align}
	\mathcal{C}_{3}[N]=&\int_{\mathcal{I}}dx\frac{(2/3)N}{4G(1+\omega{\check{Y}})}|E^{\varphi}|^{-1}|E^{x}|^{-\frac{1}{2}}\nonumber\\
	\times&(2\tilde{K}_{\varphi}E^{\varphi}+\tilde{K}_{x}E^{x})^{2},\label{sC3}\\
	\mathcal{C}_{4}[N]=\!&\int_{\mathcal{I}}\!dx\frac{2N}{3\omega}|E^{\varphi}|^{-1}\!|E^{x}|^{-\frac{1}{2}}\!(2\tilde{K}_{\varphi}E^{\varphi}\nonumber\\
	+&\tilde{K}_{x}E^{x})p^{Y},\label{sC4}
\end{align}
where $2\tilde{K}_{\varphi}E^{\varphi}+\tilde{K}_{x}E^{x}=(\tilde{K}^{1}_{\theta}E^{\theta}_{1}+\tilde{K}^{2}_{\phi}E^{\phi}_{2}+\tilde{K}^{3}_{x}E^{x}_{3})|_{\theta=\pi/2}$. The seventh term is reduced to
\begin{equation}\label{sC7}
	\mathcal{C}_{7}[N]\!=\!-\int_{\mathcal{I}}\!dx\!\frac{N}{8\rho\cdot{4}\pi}\!|E^{\varphi}||E^{x}|^{-\frac{3}{2}}(p^{x})^{2},
\end{equation}
where $|E^{\varphi}|\!|E^{x}|^{-\!3/2}\!=\!(q^{-\!3/2}\!(E^{\theta}_{1})^{2}\!(E^{\phi}_{2})^{2})\!|_{\theta=\pi/2}$. The ninth term is reduced to
\begin{equation}
	\mathcal{C}_{9}[N]\!=\!\int_{\mathcal{I}}\!dx\!N\!\sqrt{(Y_{x})^{2}\!(E^{x})^{2}\!|E^{\varphi}|^{-\!2}\!|E^{x}|^{-\!1}\!-\!\check{Y}}\partial_{x}p^{x}.
\end{equation}
where $(E^{x})^{2}|E^{\varphi}|^{-2}|E^{x}|^{-1}=q^{xx}=E^{x}_{3}E^{x}_{3}q^{-1}$. The fifth, sixth and eighth terms respectively read:
\begin{align}
	\mathcal{C}_{5}[N]\!=\!&\int_{\mathcal{I}}\!dx\frac{2NG}{3\omega^2}\!|E^{\varphi}|^{-\!1}\!|E^{x}|^{-\!\frac{1}{2}}\!(1\!+\!\omega{\check{Y}})(p^{Y})^{2},\\
	\mathcal{C}_{6}[N]=&\int_{\mathcal{I}}dx\frac{N\omega}{2G}\partial_{x}(\frac{|E^{x}|^{\frac{3}{2}}}{|E^{\varphi}|}\partial_{x}\check{Y}),\\
	\mathcal{C}_{8}[N]=&0.
\end{align}

The Poisson brackets between the reduced vector and scalar constraints are calculated as:
\begin{equation}
	\begin{split}
		&\{\mathcal{C}_{x}[N^{x}],\mathcal{C}_{x}[M^{x}]\}=\mathcal{C}_{x}[N^{x}\partial_{x}M^{x}-M^{x}\partial_{x}N^{x}],\\
		&\{\mathcal{C}_{x}[N^{x}],\mathcal{C}[M]\}=\mathcal{C}[N^{x}\partial_{x}M],\\
		&\{\mathcal{C}[N],\mathcal{C}[M]\}=\mathcal{C}_{x}[\frac{|E^{x}|}{(E^{\varphi})^{2}}(N\partial_{x}M-M\partial_{x}N)].
	\end{split}
\end{equation}
Hence they form a first-class constraint system and constitute the spherically symmetric version of the hypersurface deformation algebra.

\subsection{Solving Scalar and Vector Constraints}

The expressions of the reduced vector and scalar constraints imply that, we can choose $p^{x}$ as the spatial coordinate $\sigma$ and $\check{Y}$ as the time coordinate $\tau$ for deparametrization. It is straightforward to show that, following the procedure in subsection \ref{Deparametrization}, the reduced phase space $\bar{\Gamma}$ with all constraints resolved can be obtained, consisting of canonical pairs $(\bar{K}_{x}(\sigma,\tau)\equiv\bar{\tilde{K}}_{x}(\sigma,\tau),\bar{E}^{x}(\sigma',\tau))$ and $(\bar{K}_{\varphi}(\sigma,\tau)\equiv\bar{\tilde{K}}_{\varphi}(\sigma,\tau),\bar{E}^{\varphi}(\sigma',\tau))$, along with their nontrivial Poisson brackets:
\begin{equation}
\begin{split}
	\{\bar{K}_{x}(\sigma,\tau),\bar{E}^x(\sigma',\tau)\}&=2G\delta(\sigma,\sigma'),\\ \{\bar{K}_{\varphi}(\sigma,\tau),\bar{E}^{\varphi}(\sigma',\tau)\}&=G\delta(\sigma,\sigma').
\end{split}
\end{equation}
The physical Hamiltonian is determined by the following quadratic equation with respect to $\bar{p}^{Y}$:
\begin{equation}\label{solveH}
	\bar{\mathcal{C}}:=\mathfrak{a}(\bar{p}^{Y})^{2}+\mathfrak{b}\bar{p}^{Y}+\mathfrak{c}=0,
\end{equation}
where
\begin{equation}
\begin{split}
	\mathfrak{a}&:=\frac{2G}{3\omega^2}(1+\omega\tau)|\bar{E}^{\varphi}|^{-1}|\bar{E}^{x}|^{-1/2},\\ \mathfrak{b}&:=\frac{2}{3\omega}|\bar{E}^{\varphi}|^{-1}|\bar{E}^{x}|^{-1/2}(2\bar{K}_{\varphi}\bar{E}^{\varphi}+\bar{K}_{x}\bar{E}^{x}),\\
	\mathfrak{c}&:=\bar{\mathcal{C}}_{1}+\bar{\mathcal{C}}_{2}+\bar{\mathcal{C}}_{3}+\bar{\mathcal{C}}_{7}+\bar{\mathcal{C}}'_{9}.
\end{split}
\end{equation}
Here we defined
\begin{align}
	\bar{\mathcal{C}}_{1}&:=\frac{1+\omega\tau}{2G}|\bar{E}^{\varphi}|^{-1}|\bar{E}^{x}|^{-\frac{1}{2}}[\gamma^2(\bar{K}_{\varphi})^2(\bar{E}^{\varphi})^{2}\nonumber\\
	&+(\bar{E}^{\varphi})^{2}((\frac{\partial_{\sigma}\bar{E}^x}{2\bar{E}^{\varphi}})^{2}-1)+
	2\bar{E}^{x}\bar{E}^{\varphi}\partial_{\sigma}(\frac{\partial_{\sigma}\bar{E}^x}{2\bar{E}^{\varphi}})\nonumber\\
	&+2\gamma^2\bar{K}_{x}\bar{K}_{\varphi}\bar{E}^{x}\bar{E}^{\varphi}],\\
	\bar{\mathcal{C}}_{2}&:=\frac{-1}{2G}(\dfrac{1}{1+\omega\tau}+\gamma^2(1+\omega\tau))|\bar{E}^{\varphi}|^{-1}|\bar{E}^{x}|^{-\frac{1}{2}}\nonumber\\
	&\times\!((\bar{K}_{\varphi})^2(\bar{E}^{\varphi})^{2}+2\bar{K}_{x}\bar{K}_{\varphi}\bar{E}^{x}\bar{E}^{\varphi}),\\
	\bar{\mathcal{C}}_{3}&:=\frac{(2/3)}{4G(1+\omega\tau)}|\bar{E}^{\varphi}|^{-1}|\bar{E}^{x}|^{-\frac{1}{2}}\nonumber\\
	&\times(2\bar{K}_{\varphi}\bar{E}^{\varphi}+\bar{K}_{x}\bar{E}^{x})^{2},\\
	\bar{\mathcal{C}}_{7}&:=-\frac{1}{8\rho\cdot{4}\pi}|\bar{E}^{\varphi}||\bar{E}^{x}|^{-3/2}\sigma^{2},\\
	\bar{\mathcal{C}}'_{9}&:=\sqrt{(\bar{\mathcal{C}}^{geo}_{x})^{2}(\bar{E}^{x})^{2}|\bar{E}^{\varphi}|^{-2}|\bar{E}^{x}|^{-1}-\tau},
\end{align}
where
\begin{align}
	\bar{\mathcal{C}}^{geo}_{x}&:=\frac{1}{2G}(2\bar{E}^{\varphi}\partial_{\sigma}\bar{K}_{\varphi}-\bar{K}_{x}\partial_{\sigma}\bar{E}^{x}).
\end{align}
By the assumption that $\partial\bar{\mathcal{C}}/\partial\bar{p}^{Y}=2\mathfrak{a}\bar{p}^{Y}+\mathfrak{b}\ne0$, the phase space $\bar{\Gamma}$ can be divided into the following two regions: $\bar{\Gamma}_{+}:2\mathfrak{a}\bar{p}^{Y}+\mathfrak{b}>0$ and $\bar{\Gamma}_{-}:2\mathfrak{a}\bar{p}^{Y}+\mathfrak{b}<0$, such that
\begin{equation}
	\bar{p}^{Y}|_{\bar{\Gamma}_{\pm}}=-h_{\pm}:=\frac{-\mathfrak{b}\pm\sqrt{\mathfrak{b}^2-4\mathfrak{a}\mathfrak{c}}}{2\mathfrak{a}}.
\end{equation}
Hence, we obtain different preparatory physical Hamiltonians $H^{0}_{phy,\pm}\equiv\int_{\tilde{\mathcal{I}}}d\sigma{h}_{\pm}$ in $\bar{\Gamma}_{\pm}$.

Note that the one-dimensional manifold $\tilde{\mathcal{I}}$ could be either non-compact or bounded. In order for $H^{0}_{phy,\pm}$ to have a well-defined functional derivative, we need to add boundary terms. It turns out that, in order to cancel the boundary terms generated by the variation of $H^{0}_{phy,\pm}$, the added boundary terms $H_{bdy,\pm}$ have to satisfy the following condition:
\begin{align}
	\delta{H}_{bdy,\pm}\!=\!&\mp\!\int_{\tilde{\mathcal{I}}}\!d\sigma\partial_{\sigma}[\Delta^{-1/2}(\Lambda_{0}\delta\bar{E}^{x}\!+\!\delta\Lambda_{1}+\Lambda_{2}\nonumber\\
	\times\!&(2\bar{E}^{\varphi}\!\delta\!\bar{K}_{\varphi}\!-\!\bar{K}_{x}\!\delta\!\bar{E}^{x}))\!-\!(\partial_{\sigma}\Delta^{-\frac{1}{2}})\Lambda_{3}\delta\!\bar{E}^{x}],\label{addedboundary}
\end{align}
where
\begin{equation}
	\begin{split}
		\Delta:=&\mathfrak{b}^2-4\mathfrak{a}\mathfrak{c},\\ \Lambda_{0}:=&-\frac{1+\omega\tau}{4G}|\bar{E}^{\varphi}|^{-1}|\bar{E}^{x}|^{-\frac{1}{2}}\partial_{\sigma}\bar{E}^{x},\\
		\Lambda_{1}:=&\frac{1+\omega\tau}{2G}|\bar{E}^{\varphi}|^{-1}|\bar{E}^{x}|^{-\frac{1}{2}}\bar{E}^{x}\partial_{\sigma}\bar{E}^{x}=:\Lambda_{3}\partial_{\sigma}\bar{E}^{x},\\
		\Lambda_{2}:=&\frac{1}{2G}\bar{\mathcal{C}}^{geo}_{x}(\bar{E}^{x})^{2}|\bar{E}^{\varphi}|^{-2}|\bar{E}^{x}|^{-1}({\bar{\mathcal{C}}'_{9}})^{-1}.
	\end{split}
\end{equation}

To simplify this condition, we consider the case of asymptotically flat spacetime. By assuming $\tilde{\mathcal{I}}=\mathbb{R}$ and performing the canonical transformation:
\begin{equation}
	\begin{split}
		|\bar{E}^{x}|=&R^2,\,\qquad\bar{K}_{x}=-R^{-1}P_{R}+R^{-2}\Lambda{P}_{\Lambda},\\
		|\bar{E}^{\varphi}|=&R\Lambda,\qquad\bar{K}_{\varphi}=-R^{-1}P_{\Lambda},
	\end{split}
\end{equation}
the asymptotic flat boundary conditions as $\sigma\to\pm\infty$ read \cite{Kuchar,Giesel5}:
\begin{equation}\label{boundarycondition}
\begin{split}
		\Lambda(\sigma,\tau)=&1\!+\!M_{\pm}(\tau)|\sigma|^{-1}\!+\!O^{\infty}(|\sigma|^{-1-\varepsilon}),\\
		{P}_{\Lambda}(\sigma,\tau)=&O^{\infty}(|\sigma|^{-\varepsilon}),\\
		R(\sigma,\tau)=&|\sigma|\!+\!O^{\infty}(|\sigma|^{-\varepsilon}),\\ {P}_{R}(\sigma,\tau)=&O^{\infty}(|\sigma|^{-\varepsilon}),
\end{split}
\end{equation}
where $\varepsilon$ is a small positive number, $M_{\pm}(\tau)$ are functions that depend only on $\tau$, and the expression $f(\sigma,\tau)=O^{\infty}(|\sigma|^{-n})$ means that $f$ decays as $|\sigma|^{-n}$, $\partial_{\sigma}f$ decays as $|\sigma|^{-n-1}$, and so on for higher-order spatial derivatives. These conditions imply
\begin{equation}
	H_{bdy,\pm}\!=\!\pm\left.[\alpha^{-\frac{1}{2}}(\tau)\text{sgn}(\sigma)\frac{1+\omega\tau}{2G}|\bar{E}^{x}(\sigma)|]\right|^{+\infty}_{-\infty},
\end{equation}
where $\alpha(\tau)=-(8G/(3\omega^{2}))(1+\omega\tau)(-(8\rho\cdot4\pi)^{-1}+(-\tau)^{1/2})$. Therefore, the physical Hamiltonian in $\bar{\Gamma}_{\pm}$ read $H_{phy,\pm}\equiv\!H^{0}_{phy,\pm}+H_{bdy,\pm}$.

\section{Quantization of Reduced Phase Space}\label{QRPSVTGT}
It should be noted that, there is no reason to choose only $\bar{\Gamma}_{+}$ or $\bar{\Gamma}_{-}$ for quantization. We need to quantize both parts of the phase space, and the final physical Hilbert space should be the direct sum of the two corresponding parts. To avoid the confusion of notations, we will only demonstrate the quantization of the phase space $\bar{\Gamma}_{+}$, while the quantization of $\bar{\Gamma}_{-}$ can be achieved by using exactly the same method.

\subsection{Physical Hilbert Space}
We define a graph $\mathfrak{g}$ on $\tilde{\mathcal{I}}$ as the set consisting of a finite number of edges and points, where the points do not intersect the edges, and the edges intersect at most at their respective endpoints. In addition, we fix the orientation of each edge to be compatible with the orientation of $\tilde{\mathcal{I}}$, and define $\mathrm{Cyl}_{\mathfrak{g}}$ as the vector space generated by finite linear combinations of functions of the following form:
\begin{align}
	T_{\mathfrak{g},\vec{k},\vec{\mu}}(\bar{K}_{x},\bar{K}_{\varphi})\!=&\!\prod\limits_{e\in{E(\mathfrak{g})}}\!\exp(i\frac{k_{e}}{2}\gamma\!\int_{e}d\sigma\!\bar{K}_{x}(\sigma))\nonumber\\
	\times&\!\prod\limits_{v\in{V(\mathfrak{g})}}\!\exp(i\mu_{v}\gamma\bar{K}_{\varphi}(v)),
\end{align}
where $E(\mathfrak{g})$ is the set of all edges in $\mathfrak{g}$, $V(\mathfrak{g})$ is the set of all points with zero-valence and the endpoints of the edges in $\mathfrak{g}$, and $\mu_{v}\in\mathbb{R}$, $k_{e}\in\mathbb{Z}$ \cite{Bojowald1,Chiou,Bojowald3}. Since it is always possible to find a graph $\mathfrak{g}''$ such that $T_{\mathfrak{g},\vec{k},\vec{\mu}}$ and $T_{\mathfrak{g}',\vec{k}',\vec{\mu}'}$ can be viewed as elements of $\mathrm{Cyl}_{\mathfrak{g}''}$, we define the inner product on $\mathrm{Cyl}_{+}\equiv\cup_{\mathfrak{g}}\mathrm{Cyl}_{\mathfrak{g}}$ as:
\begin{align}
	&\big<T_{\mathfrak{g},\vec{k},\!\vec{\mu}},T_{\mathfrak{g}',\vec{k}',\vec{\mu}'}\big>_{phy,+}\nonumber\\
	=&\!\prod\limits_{e''\in{E(\mathfrak{g}'')}}\!\delta_{k_{e''},k'_{e''}}\!\prod\limits_{v''\in{V(\mathfrak{g}'')}}\!\delta_{\mu_{v''},\mu'_{v''}}.\label{scalar product6}
\end{align}
By completing $\mathrm{Cyl}_{+}$ with (\ref{scalar product6}), we get the Hilbert space $\mathcal{H}_{+}$. As a result, $\{T_{\mathfrak{g},\vec{k},\vec{\mu}}\}$ forms an orthonormal basis, where $T_{\mathfrak{g},\vec{k},\vec{\mu}}=1$ when $\mathfrak{g}=\emptyset$; otherwise, $\mu_{v}\ne0$ for each point of zero-valence, $k_{e}\ne0$ for each edge, and $\mu_{v}$ and $k_{e^{+}(v)}-k_{e^{-}(v)}$ are not both vanishing for each bivalent point, with $e^{\pm}(v)$ denoting the edges with $v$ as the beginning and final points respectively.

The actions of the operators corresponding to the conjugate momenta on the basis vector read respectively as \cite{Bojowald1,Chiou,Bojowald3}:
\begin{align}
	&\hat{\bar{E}}^{x}(\sigma)\!\cdot\! T_{\mathfrak{g},\vec{k},\vec{\mu}}\!=\!\gamma{l}^{2}_{p}\frac{k_{e^{+}(\sigma)}\!+\!k_{e^{-}(\sigma)}}{2}T_{\mathfrak{g},\vec{k},\vec{\mu}},\label{fluxcomponent1}\\
	&\int_{\tilde{\mathcal{I}}}\!d\sigma\!\lambda(\sigma)\!\hat{\bar{E}}^{\varphi}(\sigma)\!\cdot\! T_{\mathfrak{g},\vec{k},\vec{\mu}}\!=\!\gamma{l}^{2}_{p}\!\sum\limits_{v\in{V(\mathfrak{g})}}\!\lambda(v)\mu_{v}T_{\mathfrak{g},\vec{k},\vec{\mu}},\label{fluxcomponent2}
\end{align}
where ${l}^{2}_{p}=G\hbar$, and $T_{\mathfrak{g},\vec{k},\vec{\mu}}$ is adjusted to a form compatible with $\sigma$ before being acted, such that $\sigma\in V(\mathfrak{g})$, and when $\sigma$ is not a bivalent point, an edge with $k_{e}=0$ is added to make $\sigma$ a bivalent point. Then, we obtain the following useful operators:
\begin{align}
	&\int_{\tilde{\mathcal{I}}}d\sigma\lambda(\sigma)\partial_{\sigma}\hat{\bar{E}}^{x}(\sigma)\!\cdot\! T_{\mathfrak{g},\vec{k},\vec{\mu}}\nonumber\\
	=&\gamma{l}^{2}_{p}\sum\limits_{v\in{V(\mathfrak{g})}}\lambda(v)(k_{e^{+}(v)}-k_{e^{-}(v)})T_{\mathfrak{g},\vec{k},\vec{\mu}},\label{fluxcomponent3}\\
	&\hat{V}_{(\sigma,\delta)}\!\cdot\! T_{\mathfrak{g},\vec{k},\vec{\mu}}\nonumber\\
	=&\!4\pi\!\gamma^{\frac{3}{2}}l^{3}_{p}\!\sum\limits_{v\in{V(\mathfrak{g})}}\!\chi_{\delta}(\sigma,\!v)\left|\!\frac{k_{e^{+}\!(v)}\!+\!k_{e^{-}\!(v)}}{2}\!\right|^{\frac{1}{2}}\!|\mu_{v}|\!T_{\mathfrak{g},\vec{k},\vec{\mu}}\nonumber\\
	=&:\sum\limits_{v\in{V(\mathfrak{g})}}\chi_{\delta}(\sigma,v)\hat{V}_{v}\cdot T_{\mathfrak{g},\vec{k},\vec{\mu}},\label{fluxcomponent4}
\end{align}
where $V_{(\sigma,\delta)}\equiv4\pi\int_{\tilde{\mathcal{I}}}d\sigma'\chi_{\delta}(\sigma,\sigma')|\bar{E}^{x}|^{1/2}|\bar{E}^{\varphi}|(\sigma')$ is the volume smeared by the characteristic function $\chi_{\delta}(\sigma,\sigma')$. Note that Eq.(\ref{fluxcomponent3}) results from the integration by parts followed by the regularization. Similar to the full theory, we can define $\hat{V}^{-1/2}_{(\sigma,\delta)}:=\lim\limits_{\varsigma\to0}(\hat{V}_{(\sigma,\delta)}+\varsigma)^{-1}\hat{V}^{1/2}_{(\sigma,\delta)}$.

The residual gauge transformation $\Pi$ can be promoted to an unitary operator acting on the basis as \cite{Chiou,Ashtekar6}:
\begin{align}
	\hat{\Pi}\!\cdot\! T_{\mathfrak{g},\vec{k},\vec{\mu}}(\bar{K}_{x},\bar{K}_{\varphi}):=&T_{\mathfrak{g},\vec{k},\vec{\mu}}(\bar{K}_{x},-\bar{K}_{\varphi})\nonumber\\
	=&T_{\mathfrak{g},\vec{k},-\vec{\mu}}(\bar{K}_{x},\bar{K}_{\varphi}).
\end{align}
Therefore, the physical Hilbert space corresponding to $\bar{\Gamma}_{+}$ can be defined as $\mathcal{H}_{phy,+}:=\hat{P}[\mathcal{H}_{+}]$, with the projection operator $\hat{P}\equiv(1+\hat{\Pi})/2$.

\subsection{Physical Hamiltonian Operator}

To quantize the physical Hamiltonian $H_{phy,+}$, we first promote it to an operator on $\mathcal{H}_{+}$ via regularization, and then to an operator on $\mathcal{H}_{phy,+}$. For simplicity, we divide the Hamiltonian into the following three terms:
\begin{equation}
\begin{split}
	H_{1}\!&:=\!\int_{\tilde{\mathcal{I}}}d\sigma\frac{\mathfrak{b}}{2\mathfrak{a}}(\sigma),\qquad
	H_{2}\!:=\!\int_{\tilde{\mathcal{I}}}d\sigma\frac{-\Delta^{\frac{1}{2}}}{2\mathfrak{a}}(\sigma),\\
	H_{3}\!&:=\!H_{bdy,+}.
\end{split}
\end{equation}
By introducing a partition $\tilde{\mathcal{I}}=\cup_{r}\tilde{\mathcal{I}}^{\iota}_{r}$ parameterized by $\iota$, such that each cell $\tilde{\mathcal{I}}^{\iota}_{r}$ shrank uniformly as $\iota\to0$, $H_{2}$ can be regularized using techniques similar to those in Refs.\cite{Alesci:2015wla,Lewandowski2} as:
\begin{align}
	H_{2}=&\frac{-3\omega^{2}}{4G(1+\omega\tau)}\lim\limits_{\iota\to0}\sum\limits_{r}\Big[|(\int_{\tilde{\mathcal{I}}^{\iota}_{r}}d\sigma\mathfrak{B}(\sigma))^{2}\nonumber\\
	-&\frac{2G(1\!+\!\omega\tau)}{3\pi\omega^2}(\int_{\tilde{\mathcal{I}}^{\iota}_{r}}\!d\sigma4\pi|\bar{E}^{\varphi}|(\sigma)|\bar{E}^{x}|^{1/2}(\sigma))\nonumber\\
	\times&\int_{\tilde{\mathcal{I}}^{\iota}_{r}}\!d\sigma\mathfrak{c}(\sigma)|\Big]^{1/2},\label{H2_components}
\end{align}
where $\mathfrak{B}\equiv|\bar{E}^{x}|^{1/2}|\bar{E}^{\varphi}|\mathfrak{b}$. Then, we use the point-splitting scheme \cite{Thiemann4,Thiemann:1997rt,Jinsong,Lewandowski1,Alesci:2015wla,Chiou,Bojowald3} to regularize the following components in (\ref{H2_components}) as:
\begin{align}
	\mathfrak{B}[\lambda]\!&=\!\lim\limits_{\delta\to0}\!\int_{\tilde{\mathcal{I}}}\!d\sigma\!\frac{2\lambda(\sigma)}{3\omega}\!V^{-\frac{1}{2}}_{(\sigma,\delta)}(2\bar{K}_{\varphi}\!\bar{E}^{\varphi}\!+\!\bar{K}_{x}\!\bar{E}^{x})\!(\sigma)\nonumber\\
	&\times\!4\pi\!\int_{\tilde{\mathcal{I}}}\!d\sigma'\!\chi_{\delta}\!(\sigma,\!\sigma')|\bar{E}^{x}|^{\frac{1}{2}}(\sigma')|\bar{E}^{\varphi}|\!(\sigma')\!V^{-\frac{1}{2}}_{(\sigma'\!,\delta)},\label{BB}\\
	\nonumber\\
	\bar{\mathcal{C}}_{1,1}[\lambda]\!&:=\!\int_{\tilde{\mathcal{I}}}\!d\sigma\!\lambda\!\frac{1\!+\!\omega\tau}{2G}\!|\bar{E}^{\varphi}|^{-\!1}|\!\bar{E}^{x}|^{-\frac{1}{2}}\!(\gamma^2\!(\!\bar{K}_{\varphi}\!)^2\!(\!\bar{E}^{\varphi}\!)^{2}\nonumber\\
	&+2\gamma^2\bar{K}_{x}\bar{K}_{\varphi}\bar{E}^{x}\bar{E}^{\varphi}),\nonumber\\
	&=\!\lim\limits_{\delta\to0}\!\int_{\tilde{\mathcal{I}}}\!d\sigma\!2\pi\!\frac{1\!+\!\omega\tau}{G}\!\lambda(\sigma)\!V^{-\frac{1}{2}}_{(\sigma,\delta)}\!(\gamma^2\!(\bar{K}_{\varphi})^2\!\bar{E}^{\varphi}\nonumber\\
	&+\!2\gamma^2\!\bar{K}_{x}\!\bar{K}_{\varphi}\!\bar{E}^{x})\!\int_{\tilde{\mathcal{I}}}\!d\sigma'\!\chi_{\delta}(\sigma,\!\sigma')\!\bar{E}^{\varphi}\!(\sigma')\!V^{-\frac{1}{2}}_{(\sigma',\delta)},\label{C1,1}\\
	\bar{\mathcal{C}}_{1,2}[\lambda]\!&:=\!\int_{\tilde{\mathcal{I}}}\!d\sigma\!\lambda\!\frac{1\!+\!\omega\tau}{2G}\!|\bar{E}^{\varphi}|^{-\!1}\!|\bar{E}^{x}|^{-\frac{1}{2}}\!(\bar{E}^{\varphi})^{2}\nonumber\\
	&\times\!((\!\frac{\partial_{\sigma}\bar{E}^x}{2\bar{E}^{\varphi}}\!)^{2}\!-\!1),\nonumber\\
	&=\!\lim\limits_{\delta\to0}\!\int_{\tilde{\mathcal{I}}}\!d\sigma2\pi\!\frac{1\!+\!\omega\tau}{G}\!\lambda\!(\sigma)\!V^{-\frac{1}{2}}_{(\sigma,\delta)}\!(\!\frac{\partial_{x}\bar{E}^x}{2}\!+\!\bar{E}^{\varphi}\!)\!(\sigma)\nonumber\\
	&\times\int_{\tilde{\mathcal{I}}}d\sigma'\chi_{\delta}(\sigma,\sigma')(\frac{\partial_{x}\bar{E}^x}{2}-\bar{E}^{\varphi})(\sigma')V^{-\frac{1}{2}}_{(\sigma',\delta)},\label{C1,2}\\
	\bar{\mathcal{C}}_{1,3}[\lambda]\!&:=\!\int_{\tilde{\mathcal{I}}}\!d\sigma\!\lambda\!\frac{1\!+\!\omega\tau}{2G}\!|\bar{E}^{\varphi}|^{-\!1}\!|\bar{E}^{x}|^{-\frac{1}{2}}\bar{E}^{x}\!\bar{E}^{\varphi}\!\partial_{\sigma}\!(\!\frac{\partial_{\sigma}\bar{E}^x}{\bar{E}^{\varphi}}\!)\nonumber\\
	&=\!-4\pi\!\frac{1\!+\!\omega\tau}{3G}\!\lim\limits_{\delta\to0}\!\int_{\tilde{\mathcal{I}}}\!d\sigma\!\partial_{\sigma}\!(\!\lambda|\bar{E}^x|^{\frac{1}{2}}\!)\!(\sigma)\!V^{-\frac{1}{2}}_{(\sigma,\delta)}\nonumber\\
	&\times\!\int_{\tilde{\mathcal{I}}}\!d\sigma'\chi_{\delta}\!(\sigma,\sigma')\!(\partial_{\sigma'}\!|\bar{E}^x|^{\frac{3}{2}}(\sigma'))\!V^{-\frac{1}{2}}_{(\sigma',\delta)},\label{C1,3}\\
	\nonumber\\
	\bar{\mathcal{C}}_{2}[\lambda]\!&=\!\frac{-2\pi}{G}\!(\!\dfrac{1}{1\!+\!\omega\tau}\!+\!\gamma^2(1\!+\!\omega\tau)\!)\!\lim\limits_{\delta\to0}\!\int_{\tilde{\mathcal{I}}}\!d\sigma\!\lambda(\sigma)\nonumber\\
	&\times\!((\bar{K}_{\varphi})^2\!\bar{E}^{\varphi}+\!2\bar{K}_{x}\bar{K}_{\varphi}\bar{E}^{x})(\sigma)\nonumber\\
	&\times\!\int_{\tilde{\mathcal{I}}}d\sigma'\chi_{\delta}(\sigma,\sigma')\bar{E}^{\varphi}(\sigma')V^{-\frac{1}{2}}_{(\sigma',\delta)},\label{CC2}\\
	\nonumber\\
	\bar{\mathcal{C}}_{3}[\lambda]\!&=\!\lim\limits_{\delta\to0}\!\int_{\tilde{\mathcal{I}}}\!d\sigma\!\frac{(2/3)\pi\lambda(\sigma)}{G\!(1\!+\!\omega\!\tau)}V^{-\frac{1}{2}}_{(\sigma,\delta)}(2\bar{K}_{\varphi}\!\bar{E}^{\varphi}\nonumber\\
	&+\!\bar{K}_{x}\!\bar{E}^{x})(\sigma)\int_{\tilde{\mathcal{I}}}d\sigma'\chi_{\delta}(\sigma,\sigma')(2\bar{K}_{\varphi}\bar{E}^{\varphi}\nonumber\\
	&+\!\bar{K}_{x}\bar{E}^{x})(\sigma')V^{-\frac{1}{2}}_{(\sigma',\delta)},\label{CC3}\\
	\nonumber\\
	\bar{\mathcal{C}}_{7}[\lambda]&=-\frac{(4\pi)^{2}}{8\rho}\lim\limits_{\delta\to0}\int_{\tilde{\mathcal{I}}}d\sigma\lambda(\sigma)V^{-3/2}_{(\sigma,\delta)}\sigma\bar{E}^{\varphi}(\sigma)\nonumber\\
	&\times\!\int_{\tilde{\mathcal{I}}}\!d\sigma'\!\chi_{\delta}\!(\sigma,\sigma')\!\bar{E}^{\varphi}\!(\sigma')\!\int_{\tilde{\mathcal{I}}}\!d\sigma''\!\chi_{\delta}\!(\sigma',\sigma'')\sigma''\nonumber\\
	 &\times\!\bar{E}^{\varphi}\!(\sigma'')\!\int_{\tilde{\mathcal{I}}}\!d\sigma'''\!\chi_{\delta}\!(\sigma'',\sigma''')\!\bar{E}^{\varphi}\!(\sigma''')V^{-\frac{3}{2}}_{(\sigma''',\delta)},\label{CC7}\\
	\nonumber\\
	\bar{\mathcal{C}}'_{9}[\lambda]&=\lim\limits_{\iota\to0}\sum\limits_{r}\text{sgn}(\lambda(\sigma_{r}))\Big[(\int_{\tilde{\mathcal{I}}^{\iota}_{r}}d\sigma\lambda(\sigma)Q(\sigma))^{2}\nonumber\\
	&+(\int_{\tilde{\mathcal{I}}^{\iota}_{r}}d\sigma\lambda(\sigma)(-\tau)^{\frac{1}{2}})^{2}\Big]^{\frac{1}{2}},\label{SC'9}\\
	Q_{1}[\lambda]\!&:=\!\frac{1}{2G}\!\int_{\tilde{\mathcal{I}}}\!d\sigma\!\lambda2\bar{E}^{\varphi}(\!\partial_{\sigma}\!\bar{K}_{\varphi}\!)\bar{E}^{x}|\bar{E}^{\varphi}|^{-\!1}\!|\bar{E}^{x}|^{-\frac{1}{2}}\nonumber\\
	&=\!\lim\limits_{\delta\to0}\!\frac{2\pi}{G}\!\int_{\tilde{\mathcal{I}}}\!d\sigma\!\lambda(\sigma)\!V^{-\frac{1}{2}}_{(\sigma,\delta)}2(\!\partial_{\sigma}\!\bar{K}_{\varphi}\!)\!(\sigma)\!\bar{E}^{x}(\sigma)\nonumber\\
	&\times\!\int_{\tilde{\mathcal{I}}}\!d\sigma'\!\chi_{\delta}\!(\sigma,\sigma')\!\bar{E}^{\varphi}(\sigma')\!V^{-\frac{1}{2}}_{(\sigma',\delta)},\\
	Q_{2}[\lambda]\!&:=\!-\frac{1}{2G}\!\int_{\tilde{\mathcal{I}}}\!d\sigma\!\lambda\!\bar{K}_{x}(\!\partial_{\sigma}\bar{E}^{x}\!)\bar{E}^{x}\!|\bar{E}^{\varphi}|^{-\!1}|\!\bar{E}^{x}|^{-\frac{1}{2}}\nonumber\\
	&=\!\lim\limits_{\delta\to0}\!\frac{-2\pi}{G}\int_{\tilde{\mathcal{I}}}d\sigma\lambda(\sigma)V^{-\frac{1}{2}}_{(\sigma,\delta)}\bar{K}_{x}(\sigma)\bar{E}^{x}(\sigma)\nonumber\\
	&\times\!\int_{\tilde{\mathcal{I}}}\!d\sigma'\!\chi_{\delta}(\sigma,\sigma')\!(\!\partial_{\sigma'}\!\bar{E}^{x}\!)(\sigma')\!V^{-\frac{1}{2}}_{(\sigma',\delta)},
\end{align}
where $\lambda$ is a smearing function. Note that $\bar{\mathcal{C}}'_{9}[\lambda]$ is regularized in the same way as $H_{2}$ and we have $\tau<0$ due to $\check{Y}<0$. 

The operator corresponding to $\mathfrak{B}[\lambda]$ can be obtained by two steps. First, by replacing $\bar{E}^{x}(\sigma)$ and $\bar{E}^{\varphi}(\sigma)$ with the corresponding operators, its action on the basis vector reads
\begin{align}
\hat{\mathfrak{B}}[\lambda]\!\cdot\! T_{\mathfrak{g},\vec{k},\vec{\mu}}\!=\!&\lim\limits_{\delta\to0}\!\sum\limits_{v\in{V(\mathfrak{g})}}\!\frac{2}{3\omega}\Big[(\!\sum\limits_{v'\in{V(\mathfrak{g})}}\!\lambda(v')\hat{V}^{-\frac{1}{2}}_{(v',\delta)}\nonumber\\
\times\!&\chi_{\delta}(v',v)2\bar{K}_{\varphi}(v')\gamma{l}^{2}_{p}\mu_{v'})\nonumber\\
+\!&(\sum\limits_{e\in{E(\mathfrak{g})}}\!\int_{e}\!d\sigma\lambda(\sigma)\hat{V}^{-\frac{1}{2}}_{(\sigma,\delta)}\chi_{\delta}(\sigma,v)\nonumber\\
\times\!&\bar{K}_{x}\!(\sigma)\gamma{l}^{2}_{p}k_{e})\Big]\!\hat{V}_{v}\!\hat{V}^{-\frac{1}{2}}_{(v,\delta)}\!\cdot\! T_{\mathfrak{g},\vec{k},\vec{\mu}}.\label{B11}
\end{align}
By inserting $\mathrm{N}-1$ points $\sigma_{b(e)}=\sigma_{0},\sigma_{1},\cdots,\sigma_{\mathrm{N}}=\sigma_{f(e)}$, the interval $[\sigma_{b(e)},\sigma_{f(e)}=:\sigma_{b(e)}+\Delta_{e}]$ along the edge is divided into $\mathrm{N}$ segments. Then by setting $(\sigma_{n}-\sigma_{n-1})/\Delta_{e}=:\epsilon$, the integral in (\ref{B11}) can be expressed as the following Riemann sum:
\begin{align}
		&\hat{\mathfrak{B}}[\lambda]\!\cdot\! T_{\mathfrak{g},\vec{k},\vec{\mu}}\nonumber\\
		=&\!\lim\limits_{\delta\to0}\sum\limits_{v\in{V(\mathfrak{g})}}\frac{2}{3\omega}\Big[\sum\limits_{v'\in{V(\mathfrak{g})}}\lambda(v')\hat{V}^{-\frac{1}{2}}_{(v',\delta)}\chi_{\delta}(v',v)\nonumber\\
		\times&\!2\bar{K}_{\varphi}(v')\gamma{l}^{2}_{p}\mu_{v'}+\lim\limits_{\epsilon\to0}\sum\limits_{e\in{E(\mathfrak{g})}}\Big[\sum\limits_{n=1}^{\mathrm{N}-1}\epsilon\Delta_{e}\lambda(\sigma_{n-1})\nonumber\\
		\times&\!\hat{V}^{-\frac{1}{2}}_{(\sigma_{n-1},\delta)}\!\chi_{\delta}(\sigma_{n-1},\!v)\!\bar{K}_{x}(\sigma_{n-1})\!+\!\epsilon\Delta_{e}\lambda(\sigma_{\mathrm{N}})\!\hat{V}^{-\frac{1}{2}}_{(\sigma_{\mathrm{N}},\delta)}\nonumber\\
		\times&\!\chi_{\delta}(\sigma_{N},v)\bar{K}_{x}(\sigma_{\mathrm{N}})\Big]\gamma{l}^{2}_{p}k_{e}\Big]\hat{V}_{v}\hat{V}^{-\frac{1}{2}}_{(v,\delta)}\!\cdot\! T_{\mathfrak{g},\vec{k},\vec{\mu}}.\label{B111}
\end{align}
Second, we quantize $\bar{K}_{x}$ and $\bar{K}_{\varphi}$ in $\mathfrak{B}[\lambda]$, corresponding to the radial and angular components of $\bar{K}$ at the equator respectively. Note that the holonomies of $\gamma\bar{K}$ along the three directions read respectively as
\begin{equation}
	\begin{split}
		h_{\sigma,\pm}(v):=&\exp(\int_{\sigma_{v}}^{\sigma_{v}\pm\epsilon\Delta_{e^{\pm}(v)}}d\sigma\gamma\bar{K}_{x}(\sigma)\tau_3),\\
		h_{\theta}(v):=&\exp(\xi\gamma\bar{K}_{\varphi}(v)\tau_1),\\
		h_{\phi}(v):=&\exp(\xi\gamma\bar{K}_{\varphi}(v)\tau_2),
	\end{split}
\end{equation}
where $\xi$ is a small angle. They give rise to the following approximations \cite{Chiou,Bojowald3,Ashtekar6}:
\begin{align}
	\pm\!\epsilon\Delta_{e^{\pm}(v)}\gamma\bar{K}_{x}(v)\!&\approx\!-2T\!r(h_{\sigma,\pm}(v)\tau_3)\nonumber\\
	&=\!2\!\sin(\!\frac{\gamma}{2}\!\int_{\sigma_{v}}^{\sigma_{v}\!\pm\!\epsilon\Delta_{e^{\pm}(v)}}\!d\sigma\bar{K}_{x}\!(\sigma)\!),\label{approximationK1}\\
	\xi\gamma\bar{K}_{\varphi}(v)\!&\approx\!-2T\!r\!(h_{\theta}\!(v)\tau_1)\!=\!-2T\!r\!(h_{\phi}\!(v)\tau_2)\nonumber\\
	&=\!2\sin(\!\frac{1}{2}\!\xi\gamma\bar{K}_{\varphi}\!(v)\!).\label{approximationK2}
\end{align}
By using (\ref{approximationK1}) and (\ref{approximationK2}), we get the operator corresponding to $\mathfrak{B}[\lambda]$ acting on the basis as 
\begin{align}
	&\hat{\mathfrak{B}}[\lambda]\!\cdot\! T_{\mathfrak{g},\vec{k},\vec{\mu}}\!\nonumber\\
	=\!&\sum\limits_{v\in{V(\mathfrak{g})}}\frac{2\lambda(v)}{3\omega}\hat{V}^{-\frac{1}{2}}_{v}\Big[\frac{2}{i\xi\gamma}\bigg(\exp(\frac{i\xi\gamma}{2}\bar{K}_{\varphi}(v))\nonumber\\
	-&\exp(-\frac{i}{2}\xi\gamma\bar{K}_{\varphi}(v))\bigg)
	\gamma{l}^{2}_{p}\mu_{v}\nonumber\\
	+&\sum\limits_{\varrho=\pm1}\frac{\varrho}{i}
	\Bigg(\exp(\frac{i}{2}\int_{\sigma_{v}}^{\sigma_{v}+\varrho\epsilon\Delta_{e^{\varrho}(v)}}d\sigma\gamma\bar{K}_{x}(\sigma))\nonumber\\
	-&\exp(-\frac{i}{2}\int_{\sigma_{v}}^{\sigma_{v}+\varrho\epsilon\Delta_{e^{\varrho}(v)}}d\sigma\gamma\bar{K}_{x}(\sigma))\Bigg)\nonumber\\
	\times&{l}^{2}_{p}k_{e^{\varrho}(v)}\Big]\hat{V}^{\frac{1}{2}}_{v}\cdot T_{\mathfrak{g},\vec{k},\vec{\mu}}\nonumber\\
	=:&\sum\limits_{v\in{V(\mathfrak{g})}}\lambda(v)\hat{\mathfrak{B}}_{v}\cdot T_{\mathfrak{g},\vec{k},\vec{\mu}}.
\end{align}
For simplicity, we treat $\epsilon$ and $\xi$ as two small constants in the expressions of the holonomies.

Since the structures of $H_{1}$, $\bar{\mathcal{C}}_{1,2}[\lambda]$, $\bar{\mathcal{C}}_{2}[\lambda]$, $\bar{\mathcal{C}}_{3}[\lambda]$, $\bar{\mathcal{C}}_{7}[\lambda]$, and $Q_{2}[\lambda]$ are similar to that of $\mathfrak{B}[\lambda]$. By repeating a similar construction, we obtain their corresponding operators acting on the basis respectively as:
\begin{align}
	\hat{H}_{1}\!\cdot\! T_{\mathfrak{g},\vec{k},\vec{\mu}}=\hat{\mathfrak{B}}[\frac{3\omega^{2}}{4G(1+\omega\tau)}]\!\cdot\!T_{\mathfrak{g},\vec{k},\vec{\mu}},
\end{align}
\begin{align}
	&\hat{\bar{\mathcal{C}}}_{1,2}[\lambda]\!\cdot\!T_{\mathfrak{g},\vec{k},\vec{\mu}}\nonumber\\
	=&2\pi\!\frac{1\!+\!\omega\tau}{G}\!\sum\limits_{v\in{V(\mathfrak{g})}}\!\lambda(v)\!\gamma^{2}{l}^{4}_{p}\hat{V}^{-\frac{1}{2}}_{v}\!(\frac{(k_{e^{+}\!(v)}\!-\!k_{e^{-}\!(v)}\!)^{2}}{4}\nonumber\\
	-&\!\mu_{v}^{2})\hat{V}^{-\frac{1}{2}}_{v}\!\cdot\! T_{\mathfrak{g},\vec{k},\vec{\mu}},\\
	\nonumber\\
	&\hat{\bar{\mathcal{C}}}_{2}[\lambda]\cdot T_{\mathfrak{g},\vec{k},\vec{\mu}}\nonumber\\
	=&\frac{-2\pi}{G}(\dfrac{1}{1+\omega\tau}+\gamma^2(1+\omega\tau))\sum\limits_{v\in{V(\mathfrak{g})}}\lambda(v)\nonumber\\
	\times&\!\Big[\!\frac{-{l}^{2}_{p}\mu_{v}}{\xi^{2}\gamma}\!\left(\!\exp(\frac{i}{2}\xi\!\gamma\!\bar{K}_{\varphi}\!(v))
	\!-\!\exp(\!-\frac{i}{2}\xi\!\gamma\!\bar{K}_{\varphi}\!(v)\!)\!\right)^{2}\nonumber\\
	+&\!\frac{2}{i\xi\!\gamma}\!\left(\!\exp(\frac{i}{2}\xi\!\gamma\!\bar{K}_{\varphi}\!(v))\!-\!\exp(-\frac{i}{2}\xi\!\gamma\!\bar{K}_{\varphi}\!(v))\!\right)\nonumber\\
	\times&\sum\limits_{\varrho=\pm1}\frac{\varrho}{i}\Bigg(\exp(\frac{i}{2}\int_{\sigma_{v}}^{\sigma_{v}+\varrho\epsilon\Delta_{e^{\varrho}(v)}}d\sigma\gamma\bar{K}_{x}(\sigma))\nonumber\\
	-&\!\exp(\!-\frac{i}{2}\!\int_{\sigma_{v}}^{\sigma_{v}+\varrho\epsilon\Delta_{e^{\varrho}(v)}}\!d\sigma\!\gamma\bar{K}_{x}\!(\sigma)\!)\Bigg){l}^{2}_{p}k_{e^{\varrho}(v)}\Big]\nonumber\\
	\times&\gamma{l}^{2}_{p}\mu_{v}\hat{V}^{-\frac{1}{2}}_{v}\cdot T_{\mathfrak{g},\vec{k},\vec{\mu}},\\
	\nonumber\\
	&\hat{\bar{\mathcal{C}}}_{7}[\lambda]\!\cdot \!T_{\mathfrak{g},\vec{k},\vec{\mu}}\nonumber\\
	=&\!-\frac{(4\pi)^{2}}{8\rho}\!\sum\limits_{v\in{V(g)}}\!\lambda(v)\!\sigma_{v}^{2}\hat{V}^{-\frac{3}{2}}_{v}\!(\!\gamma{l}^{2}_{p}\mu_{v}\!)^{4}\!\hat{V}^{-\frac{3}{2}}_{v}\!\cdot\!T_{\mathfrak{g},\vec{k},\vec{\mu}},\\
	\nonumber\\
	&\hat{Q}_{2}[\lambda]\!\cdot\!T_{\mathfrak{g},\vec{k},\vec{\mu}}\nonumber\\
	=&\frac{-2\pi}{G}\sum\limits_{v\in{V(\mathfrak{g})}}\lambda(v)\hat{V}^{-\frac{1}{2}}_{v}\nonumber\\
	\times&\Big[\sum\limits_{\varrho=\pm1}\frac{\varrho}{i}\Bigg(\exp(\frac{i}{2}\int_{\sigma_{v}}^{\sigma_{v}+\varrho\epsilon\Delta_{e^{\varrho}(v)}}d\sigma\gamma\bar{K}_{x}(\sigma))\nonumber\\
	-&\!\exp(-\frac{i}{2}\int_{\sigma_{v}}^{\sigma_{v}\!+\!\varrho\epsilon\Delta_{e^{\varrho}(v)}}\!d\sigma\gamma\!\bar{K}_{x}\!(\sigma))\Bigg){l}^{2}_{p}k_{e^{\varrho}(v)}\Big]\nonumber\\
	\times&\gamma{l}^{2}_{p}(k_{e^{+}(v)}-k_{e^{-}(v)})\hat{V}^{-\frac{1}{2}}_{v}\!\cdot\! T_{\mathfrak{g},\vec{k},\vec{\mu}},\\
	\nonumber\\
	&\hat{\bar{\mathcal{C}}}_{3}[\lambda]\!\cdot\!T_{\mathfrak{g},\vec{k},\vec{\mu}}\nonumber\\
	=&\!\sum\limits_{v\in{V(\mathfrak{g})}}\frac{(2/3)\pi\lambda(v)}{G(1+\omega\tau)}\hat{V}^{-\frac{1}{2}}_{v}\hat{\bar{K}}_{v}^{2}\hat{V}^{-\frac{1}{2}}_{v}\!\cdot\!T_{\mathfrak{g},\vec{k},\vec{\mu}},
\end{align}
with
\begin{align}
	&\hat{\bar{K}}_{v}\!\cdot\!T_{\mathfrak{g},\vec{k},\vec{\mu}}\nonumber\\
	:=&\Big[\!\frac{2}{i\xi}\!\left(\!\exp(\!\frac{i}{2}\xi\!\gamma\!\bar{K}_{\varphi}\!(v))\!-\!\exp(\!-\frac{i}{2}\!\xi\!\gamma\!\bar{K}_{\varphi}\!(v))\!\right)\!{l}^{2}_{p}\mu_{v}\nonumber\\
	+&\sum\limits_{\varrho=\pm1}\frac{\varrho}{i}\Bigg(\exp(\frac{i}{2}\int_{\sigma_{v}}^{\sigma_{v}+\varrho\epsilon\Delta_{e^{\varrho}(v)}}d\sigma\gamma\bar{K}_{x}(\sigma))\nonumber\\
	-&\exp(-\frac{i}{2}\int_{\sigma_{v}}^{\sigma_{v}+\varrho\epsilon\Delta_{e^{\varrho}(v)}}d\sigma\gamma\bar{K}_{x}(\sigma))\Bigg)\nonumber\\
	\times&{l}^{2}_{p}k_{e^{\varrho}(v)}\Big]\!\cdot\!T_{\mathfrak{g},\vec{k},\vec{\mu}}.
\end{align}

In the expressions of $\bar{\mathcal{C}}_{1,1}[\lambda]$ and $Q_{1}[\lambda]$, the $\bar{K}_{x}$ and $\bar{K}_{\varphi}$ come from the components of the curvature $\tilde{F}$. Therefore, for their regularization we introduce the following holonomies of loops formed by small radial and angular segments:
\begin{equation}
	\begin{split}
		h_{\alpha_{\theta\phi}}(v)\!:=\!&h_{\theta}(v)h_{\phi}(v)(h_{\theta}(v))^{-\!1}(h_{\phi}(v))^{-\!1},\\
		h_{\alpha_{\phi\sigma,\pm}}(v)\!:=\!&h_{\phi}(v)h_{\sigma,\pm}(v)(h_{\phi}(v_{\pm}))^{-\!1}(h_{\sigma,\pm}(v))^{-\!1},\\
		h_{\alpha_{\sigma\theta,\pm}}(v)\!:=\!&h_{\sigma,\pm}(v)h_{\theta}(v_{\pm})(h_{\sigma,\pm}(v))^{-\!1}(h_{\theta}(v))^{-\!1},
	\end{split}
\end{equation}
where $v_{\pm}$ is the point satisfying $\sigma_{v_{\pm}}=\sigma_{v}\pm\epsilon\Delta_{e^{\pm}(v)}$. They give rise to the following approximations \cite{Bojowald3,Ashtekar6}:
\begin{align}
	&\xi^{2}\gamma^2(\bar{K}_{\varphi})^2(v)\nonumber\\
	\approx&\!-2T\!r(h_{\alpha_{\theta\phi}}(v)\tau_3)\!=\!\sin^{2}(\xi\gamma\bar{K}_{\varphi}(v)),\label{TP}\\
	\nonumber\\
	&\pm\epsilon\Delta_{e^{\pm}(v)}\xi\gamma^2\bar{K}_{x}(v)\bar{K}_{\varphi}(v)\nonumber\\
	\approx&\!-2T\!r(h_{\alpha_{\phi\sigma,\pm}}(v)\tau_1)\!=\!-2T\!r(h_{\alpha_{\sigma\theta,\pm}}(v)\tau_2)\nonumber\\
	=&\!2\cos(\frac{1}{2}\xi\gamma\bar{K}_{\varphi}(v))\sin(\frac{1}{2}\xi\gamma\bar{K}_{\varphi}(v_{\pm}))\nonumber\\
	\times&\!\sin(\int_{\sigma_{v}}^{\sigma_{v}\pm\epsilon\Delta_{e^{\pm}(v)}}d\sigma\gamma\bar{K}_{x}(\sigma)),\label{XP}\\
	\nonumber\\
	&\pm\epsilon\Delta_{e^{\pm}(v)}\xi\gamma\partial_{\sigma}\bar{K}_{\varphi}(v)\nonumber\\
	\approx&\!-\!2T\!r(h_{\alpha_{\sigma\theta,\pm}}(v)\tau_1)\!=\!-2T\!r((h_{\alpha_{\phi\sigma,\pm}}(v))^{\!-\!1}\tau_2)\nonumber\\
	=&\!-2\Big[\sin(\frac{1}{2}\xi\gamma\bar{K}_{\varphi}(v))\cos(\frac{1}{2}\xi\gamma\bar{K}_{\varphi}(v_{\pm}))\nonumber\\
	-&\!\cos(\frac{1}{2}\xi\gamma\bar{K}_{\varphi}(v))\sin(\frac{1}{2}\xi\gamma\bar{K}_{\varphi}(v_{\pm}))\nonumber\\
	\times&\!\cos(\int_{\sigma_{v}}^{\sigma_{v}\pm\epsilon\Delta_{e^{\pm}(v)}}d\sigma\gamma\bar{K}_{x}(\sigma))\Big].\label{XP2}
\end{align}
By using expressions of (\ref{TP}), (\ref{XP}) and (\ref{XP2}), the operators corresponding to $\bar{\mathcal{C}}_{1,1}[\lambda]$ and $Q_{1}[\lambda]$ are given by:
\begin{align}
	&\hat{\bar{\mathcal{C}}}_{1,1}[\lambda]\!\cdot\! T_{\mathfrak{g},\vec{k},\vec{\mu}}\nonumber\\
	=&\sum\limits_{v\in{V(\mathfrak{g})}}2\pi\frac{1+\omega\tau}{G}\lambda(v)\hat{V}^{-\frac{1}{2}}_{v}\nonumber\\
	\times&\!\Big[\!\frac{-1}{4\xi^{2}}\big(\exp(i\xi\!\gamma\!\bar{K}_{\varphi}\!(v))\!-\!\exp(-i\xi\!\gamma\!\bar{K}_{\varphi}\!(v))\big)^{2}\gamma{l}^{2}_{p}\mu_{v}\nonumber\\
	+&\!\sum\limits_{\varrho=\pm1}\!\frac{\varrho}{i\xi}\!\left(\!\exp(\frac{i}{2}\xi\!\gamma\!\bar{K}_{\varphi}\!(v))\!+\!\exp(\!-\frac{i}{2}\xi\!\gamma\!\bar{K}_{\varphi}\!(v))\!\right)\nonumber\\
	\times&\left(\exp(\frac{i}{2}\xi\gamma\bar{K}_{\varphi}(v_{\varrho}))-\exp(-\frac{i}{2}\xi\gamma\bar{K}_{\varphi}(v_{\varrho}))\right)\nonumber\\
	\times&\frac{1}{2i}\Bigg(\exp(i\int_{\sigma_{v}}^{\sigma_{v}+\varrho\epsilon\Delta_{e^{\varrho}(v)}}d\sigma\gamma\bar{K}_{x}(\sigma))\nonumber\\
	-&\!\exp(\!-i\int_{\sigma_{v}}^{\sigma_{v}\!+\!\varrho\epsilon\Delta_{e^{\varrho}(v)}}d\sigma\gamma\bar{K}_{x}(\sigma))\Bigg)\gamma{l}^{2}_{p}k_{e^{\varrho}(v)}\Big]\nonumber\\
	\times&\gamma{l}^{2}_{p}\mu_{v}\hat{V}^{-\frac{1}{2}}_{v}\!\cdot\! T_{\mathfrak{g},\vec{k},\vec{\mu}},\\
	\nonumber\\
	&\hat{Q}_{1}[\lambda]\!\cdot\! T_{\mathfrak{g},\vec{k},\vec{\mu}}\nonumber\\
	=&\frac{2\pi}{G}\sum\limits_{v\in{V(g)}}\lambda(v)\hat{V}^{-\frac{1}{2}}_{v}\frac{-4}{\xi\gamma}\Big[\sum\limits_{\varrho=\pm1}\frac{\varrho}{4i}\nonumber\\
	\times&\Big[\left(\exp(\frac{i}{2}\xi\gamma\bar{K}_{\varphi}(v))-\exp(-\frac{i}{2}\xi\gamma\bar{K}_{\varphi}(v))\right)\nonumber\\
	\times&\left(\exp(\frac{i}{2}\xi\gamma\bar{K}_{\varphi}(v_{\varrho}))+\exp(-\frac{i}{2}\xi\gamma\bar{K}_{\varphi}(v_{\varrho}))\right)\nonumber\\
	-&\left(\exp(\frac{i}{2}\xi\gamma\bar{K}_{\varphi}(v))+\exp(-\frac{i}{2}\xi\gamma\bar{K}_{\varphi}(v))\right)\nonumber\\
	\times&\left(\exp(\frac{i}{2}\xi\gamma\bar{K}_{\varphi}(v_{\varrho}))-\exp(-\frac{i}{2}\xi\gamma\bar{K}_{\varphi}(v_{\varrho}))\right)\nonumber\\
	\times&\frac{1}{2}\Bigg(\exp(i\int_{\sigma_{v}}^{\sigma_{v}+\varrho\epsilon\Delta_{e^{\varrho}(v)}}d\sigma\gamma\bar{K}_{x}(\sigma))\nonumber\\
	+&\!\exp(\!-i\int_{\sigma_{v}}^{\sigma_{v}\!+\!\varrho\epsilon\Delta_{e^{\varrho}(v)}}\!d\sigma\gamma\bar{K}_{x}\!(\sigma))\Bigg)\Big]\gamma{l}^{2}_{p}k_{e^{\varrho}(v)}\Big]\nonumber\\
	\times&\gamma{l}^{2}_{p}\mu_{v}\hat{V}^{-\frac{1}{2}}_{v}\cdot T_{\mathfrak{g},\vec{k},\vec{\mu}}.
\end{align}

To quantize $\bar{\mathcal{C}}_{1,3}[\lambda]$, we first replace the second integral in (\ref{C1,3}) with the corresponding operator and apply it to the basis vector as:
\begin{align}
		&\hat{\bar{\mathcal{C}}}_{1,3}[\lambda]\!\cdot\! T_{\mathfrak{g},\vec{k},\vec{\mu}}\nonumber\\
		=&-4\pi\frac{(1+\omega\tau)}{3G}\lim\limits_{\delta\to0}\sum\limits_{v\in{V(\mathfrak{g})}}\int_{\tilde{\mathcal{I}}}d\sigma\partial_{\sigma}(\lambda|\bar{E}^x|^{\frac{1}{2}})(\sigma)\nonumber\\
		\times&V^{-\frac{1}{2}}_{(\sigma,\delta)}\chi_{\delta}(\sigma,v)\gamma^{3/2}{l}^{3}_{p}(|k_{e^{+}(v)}|^{\frac{3}{2}}-|k_{e^{-}(v)}|^{3/2})\nonumber\\
		\times& V^{-\frac{1}{2}}_{(v,\delta)}\!\cdot\!T_{\mathfrak{g},\vec{k},\vec{\mu}}.
\end{align}
Then, by performing integration by parts and replacing $\bar{E}^{x}(\sigma)$ with $\hat{\bar{E}}^{x}(\sigma)$, the result is the sum of integrals along all edges $e\in E(\mathfrak{g})$. Approximating the integrals by Riemann sums, we obtain
\begin{align}
		&\hat{\bar{\mathcal{C}}}_{1,3}[\lambda]\!\cdot\! T_{\mathfrak{g},\vec{k},\vec{\mu}}\nonumber\\
		=&\!4\pi\frac{1\!+\!\omega\tau}{3G}\!\lim\limits_{\delta\to0}\!\sum\limits_{v\in{V(\mathfrak{g})}}\!\lim\limits_{\epsilon\to0}\!\sum\limits_{e\in{E(\mathfrak{g})}}\![\!\sum\limits_{n=1}^{\mathrm{N}-1}\epsilon\Delta_{e}\!\lambda(\sigma_{n-1})\nonumber\\
		\times&\partial_{\sigma}(V^{-1/2}_{(\sigma,\delta)}\chi_{\delta}(\sigma,v))(\sigma_{n-1})|k_{e}|^{1/2}\nonumber\\
		+&\epsilon\Delta_{e}\lambda(\sigma_{\mathrm{N}})\partial_{\sigma}(V^{-1/2}_{(\sigma,\delta)}\chi_{\delta}(\sigma,v))(\sigma_{\mathrm{N}-1})|k_{e}|^{1/2}]\nonumber\\
		\times&\gamma^{2}{l}^{4}_{p}(|k_{e^{+}(v)}|^{3/2}-|k_{e^{-}(v)}|^{3/2})V^{-1/2}_{(v,\delta)}\cdot T_{\mathfrak{g},\vec{k},\vec{\mu}}.\label{regularizeCC13}
\end{align}
Using the approximation 
\begin{align}
	&\epsilon\Delta_{e}\partial_{\sigma}(V^{-1/2}_{(\sigma,\delta)}\chi_{\delta}(\sigma,\!v))(\sigma_{n\!-\!1})\nonumber\\
	\approx&\!{V}^{-1/2}_{(\sigma_{n},\delta)}\chi_{\delta}(\sigma_{n},\!v)\!-\!V^{-1/2}_{(\sigma_{n\!-\!1},\delta)}\chi_{\delta}(\sigma_{n\!-\!1},\!v),
\end{align}
and replacing $V^{-1/2}_{(\sigma,\delta)}$ by its operator, we obtain the operator corresponding to $\bar{\mathcal{C}}_{1,3}[\lambda]$ acting on the basis as:
\begin{align}
		&\hat{\bar{\mathcal{C}}}_{1,3}[\lambda]\!\cdot\! T_{\mathfrak{g},\vec{k},\vec{\mu}}\nonumber\\
		=&\!4\pi\!\frac{1\!+\!\omega\tau}{3G}\!\sum\limits_{v\in{V(\mathfrak{g})}}\!\lambda(v)\hat{V}^{-\frac{1}{2}}_{v}\!(|k_{e^{-}\!(v)}|^{\frac{1}{2}}\!-\!|k_{e^{+}\!(v)}|^{\frac{1}{2}})\nonumber\\
		\times&\gamma^{2}{l}^{4}_{p}(|k_{e^{+}(v)}|^{\frac{3}{2}}\!-\!|k_{e^{-}(v)}|^{\frac{3}{2}})\hat{V}^{-\frac{1}{2}}_{v}\!\cdot\!T_{\mathfrak{g},\vec{k},\vec{\mu}}.
\end{align}

Notice that, by construction, we have
\begin{align}
	\hat{Q}[\lambda]\!\cdot\! T_{\mathfrak{g},\vec{k},\vec{\mu}}\!:=&\!(\hat{Q}_{1}[\lambda]+\hat{Q}_{2}[\lambda])\cdot  T_{\mathfrak{g},\vec{k},\vec{\mu}}\nonumber\\
	=:&\!\sum\limits_{v\in{V(\mathfrak{g})}}\lambda(v)\hat{Q}_{v}\!\cdot\! T_{\mathfrak{g},\vec{k},\vec{\mu}}.
\end{align}
Hence the operator $\hat{Q}[\lambda]$ acts only on the vertices. Hence for sufficiently small $\iota$, the action of the operator corresponding to $\int_{\tilde{\mathcal{I}}^{\iota}_{r}}d\sigma\lambda(\sigma)Q(\sigma)$ in (\ref{SC'9}) on a basis vector will depend on whether $\tilde{\mathcal{I}}^{\iota}_{r}$ contains a vertex of the quantum state. Analogous to Ref.\cite{Lewandowski2}, if $\tilde{\mathcal{I}}^{\iota}_{r}$ does not contain any vertex, only the term $\int_{\tilde{\mathcal{I}}^{\iota}_{r}}d\sigma\lambda(\sigma)(-\tau)^{1/2}$ contributes nontrivial result, while if it contains a vertex, only the term $\int_{\tilde{\mathcal{I}}^{\iota}_{r}}d\sigma\lambda(\sigma)\hat{Q}(\sigma)$ contributes. Hence, $\bar{\mathcal{C}}'_{9}[\lambda]$ can be promoted to the following operator acting on the basis as:
\begin{align}
	&\hat{\bar{\mathcal{C}}}'_{9}[\lambda]\!\cdot\!T_{\mathfrak{g},\vec{k},\vec{\mu}}\nonumber\\
	=&\!\Big(\sum\limits_{v\in{V(\mathfrak{g})}}\!\lambda(v)|\hat{Q}_{v}|\!+\!\int_{\tilde{\mathcal{I}}}\!d\sigma\!\lambda(\sigma)(-\tau)^{\frac{1}{2}}\Big)\!\cdot\!T_{\mathfrak{g},\vec{k},\vec{\mu}},
\end{align}
where $|\hat{Q}_{v}|\equiv\sqrt{\hat{Q}_{v}^{\dagger}\hat{Q}_{v}}$. 

Collecting the above results, we obtain the operator corresponding to $\mathfrak{c}[\lambda]$ acting on the basis as:
\begin{align}
	&\hat{\mathfrak{c}}[\lambda]\!\cdot\!T_{\mathfrak{g},\vec{k},\vec{\mu}}\nonumber\\
	=&(\sum\limits_{i=1}^{3}\hat{\bar{\mathcal{C}}}_{i}[\lambda]+\hat{\bar{\mathcal{C}}}_{7}[\lambda]+\hat{\bar{\mathcal{C}}}'_{9}[\lambda])\!\cdot\!T_{\mathfrak{g},\vec{k},\vec{\mu}}\nonumber\\
	=:&\!(\!\sum\limits_{v\in{V(\mathfrak{g})}}\lambda(v)\hat{\mathfrak{c}}_{v}\!+\!\int_{\tilde{\mathcal{I}}}d\sigma\lambda(\sigma)(-\tau)^{\frac{1}{2}})\!\cdot\!T_{\mathfrak{g},\vec{k},\vec{\mu}}, 
\end{align}
where $\hat{\bar{\mathcal{C}}}_{1}[\lambda]\equiv\sum\limits_{i=1}^{3}\hat{\bar{\mathcal{C}}}_{1,i}[\lambda]$. For small enough $\iota$, by replacing the variables in (\ref{H2_components}) with the corresponding operators, the operator corresponding to $H_2$ only acts on the vertices of a quantum state due to the property of $\hat{V}$. Therefore, its action on the basis reads
\begin{align}
	&\hat{H}_{2}\!\cdot\! T_{\mathfrak{g},\vec{k},\vec{\mu}}\nonumber\\
	=&\!\frac{-3\omega^{2}}{4G\!(\!1\!+\!\omega\tau\!)}\!\sum\limits_{v\in{V(\mathfrak{g})}}\!\Big|\!(\!\hat{\mathfrak{B}}_{v}\!)^{2}\!-\!\frac{2G\!(\!1\!+\!\omega\tau\!)}{3\pi\omega^2}\!\hat{V}_{v}\hat{\mathfrak{c}}_{v}\!\Big|^{\frac{1}{2}}\!\cdot\! T_{\mathfrak{g},\vec{k},\vec{\mu}}\nonumber\\
	=:&\!\frac{-3\omega^{2}}{4G(1\!+\!\omega\tau)}\!\sum\limits_{v\in{V(\mathfrak{g})}}|\hat{H}_{v}|^{\frac{1}{2}}\!\cdot\! T_{\mathfrak{g},\vec{k},\vec{\mu}},
\end{align}
where $|\hat{H}_{v}|\equiv\sqrt{\hat{H}_{v}^{\dagger}\hat{H}_{v}}$.

Similar to the treatment in Ref.\cite{Thiemann:1997rs}, the boundary term $H_{3}$ can also be promoted to an operator acting on the basis as:
\begin{align}
	&\hat{H}_{3}\!\cdot\! T_{\mathfrak{g},\vec{k},\vec{\mu}}\nonumber\\
	=&\alpha^{\!-\frac{1}{2}}\!(\tau)\!\frac{1\!+\!\omega\tau}{2G}\!\lim_{\sigma\to\infty}(|\hat{\bar{E}}^{x}\!(\sigma)|\!+\!|\hat{\bar{E}}^{x}\!(-\sigma)|)\!\cdot\! T_{\mathfrak{g},\vec{k},\vec{\mu}}.
\end{align}

Collecting the above results, we obtain the operator $\hat{H}_{+}\equiv\hat{H}_{1}+\hat{H}_{2}+\hat{H}_{3}$ on the Hilbert space $\mathcal{H}_{+}$. Since the classical physical Hamiltonian is invariant under $\Pi$-reflection, we define the physical Hamiltonian operator on the physical Hilbert space $\mathcal{H}_{phy,+}$ as $\hat{H}_{phy,+}:=\hat{P}\hat{H}_{+}\hat{P}$. Furthermore, to ensure that the evolution is unitary, we define the symmetric physical Hamiltonian operator $\hat{H}^{sym}_{phy,+}:=(\hat{H}_{phy,+}+\hat{H}^{\dagger}_{phy,+})/2$ and assume that it has a natural self-adjoint extension. At last, the final physical Hilbert space is $\mathcal{H}_{phy}\equiv\mathcal{H}_{phy,+}\oplus\mathcal{H}_{phy,-}$, and the final physical Hamiltonian operator is $\hat{H}_{phy}^{sym}\equiv\hat{H}_{phy,+}^{sym}\oplus\hat{H}_{phy,-}^{sym}$.

\section{Conclusions}\label{CR}

To summarize, the loop quantization of the vector-tensor theory has been carried out in section \ref{LQVTGT}. To this end, we started with a Hamiltonian analysis of the theory to obtain its geometric dynamics. Then, the connection dynamics of the theory was obtained through canonical transformations and phase space extension. In this formulation, the Gauss, vector, and scalar constraints form a first-class constraint system. By defining the cylindrical functions corresponding to the $\mathrm{SU}(2)$-connection $A^{i}_{a}$, vector field $Y_{a}$, and scalar field $\check{Y}$ respectively, as well as the inner product between these functions, we constructed the kinematical Hilbert space $\mathcal{H}_{kin}$ for the quantum theory. The Gauss and vector constraints have been solved at the quantum level, resulting in the $\mathrm{SU}(2)$ gauge-invariant and diffeomorphism-invariant Hilbert space $\mathcal{H}^{G}_{Diff}$. The scalar constraint was first regularized appropriately to become the operators $\hat{C}^{\epsilon}[N]$ on $\mathcal{H}_{kin}$, and then was promoted to the operators $\hat{C}^{*}[N]$ on the vertex Hilbert space $\mathcal{H}_{vtx}$ where the regularization parameter $\epsilon$ can be removed. The remaining issue of the loop quantum vector-tensor gravity, which we obtained, is how to solve the kernel of scalar constraint operator $\hat{C}^{*}[N]$, which deserve further investigations. The same open issue remains in the LQG of GR \cite{Ashtekar1,Thiemann,Rovelli:2004tv,Muxin,Thiemann:1996av}.

One way to understand the physical solutions of an either classical or quantum theory of gravity is the deparametrization scheme. For simplicity, the deparametrization of the vector-tensor theory in the spherically symmetric model has also been studied in section \ref{DVTGSSM}, followed by the corresponding quantization of the reduced phase space in section \ref{QRPSVTGT}. Following the general procedure of deparametrizing theories with diffeomorphism invariance using spacetime scalar fields, we performed deparametrization in the spherically symmetric model by utilizing the degrees of freedom of the vector field as the radial spatial and time coordinates. As a result, we obtained the reduced phase space $\bar{\Gamma}_{\pm}$ with all constraints solved, along with the time-dependent physical Hamiltonian $H_{phy,\pm}$ that generates the evolution relative to a physical time. Based on the deparametrized model, we performed loop quantization on the reduced phase space. The physical Hilbert spaces $\mathcal{H}_{phy,\pm}$ were obtained, and the physical Hamiltonian was promoted to well-defined operators on them, thereby bypassing the difficulty of solving the scalar constraint at the quantum level \cite{Domagala:2010bm,Giesel1,Giesel2,Giesel6,Thiemann3,Giesel4}.

The above results lay the foundation for further exploration of the quantum dynamics and effective dynamics of vector-tensor theory in the future. Just as LQG has been applied to cosmology \cite{Ashtekar:2003hd,Ashtekar:2011ni} and black hole physics \cite{Ashtekar6,Lewandowski:2022zce}, the loop quantum vector-tensor theory developed in this paper can similarly be applied to these areas. Of course, some open issues in LQG, such as the analysis of operator self-adjointness \cite{Thiemann:1996av,Zhang:2018wbc} and the comparison between Dirac quantization and reduced phase space quantization \cite{Giesel4,Giesel:2017mfc}, also appear in this theory and remain to be resolved.

\begin{acknowledgments}
	
This work is supported by the National Natural Science Foundation of China (Grant Nos. 12275022). We thank Cong Zhang, Haida Li, and Faqiang Yuan for helpful discussions.
	
\end{acknowledgments}

\onecolumngrid
\appendix
\section{Derivation of the Constraint Algebra}\label{appendix}
Note that the Poisson brackets between the vector constraint and the fundamental variables are given by:
\begin{align}
\{\check{Y}(x),\vec{C}[\vec{N}]\}&=\mathscr{L}_{\vec{N}}\check{Y}(x),&
\{\pi^{Y}(x),\vec{C}[\vec{N}]\}&=\mathscr{L}_{\vec{N}}\pi^{Y}(x),\\
\{Y_{a}(x),\vec{C}[\vec{N}]\}&=\mathscr{L}_{\vec{N}}Y_{a}(x),&
\{\pi^{b}(x),\vec{C}[\vec{N}]\}&=\mathscr{L}_{\vec{N}}\pi^{b}(x),\\
\{q_{ab}(x),\vec{C}[\vec{N}]\}&=\mathscr{L}_{\vec{N}}q_{ab}(x),&
\{p^{cd}(x),\vec{C}[\vec{N}]\}&=\mathscr{L}_{\vec{N}}p^{cd}(x).
\end{align}
It follows that the first two equalities in (\ref{constraint algebra}) are satisfied automatically. 

To compute the Poisson bracket between two scalar constraints, we decompose it into three separate parts. The first part is given by $\mathcal{F}_{1}\equiv\int_{\Sigma}d^3x(\delta\!C[N]/{\delta\!\check{Y}(x)})(\delta\!C[M]/{\delta\!\pi^{Y}(x)})-(\delta\!C[N]/{\delta\!\pi^{Y}(x)})(\delta\!C[M]/{\delta\!\check{Y}(x)})$, and its individual terms are evaluated as follows:
\begin{align}
\frac{\delta\!C[N]}{\delta\!\check{Y}(x)}&=\frac{\omega q^{1/2}}{8\pi G}D_{a}D^{a}N+N\mathfrak{X}_1,\\
\frac{\delta\!C[N]}{\delta\!\pi^{Y}(x)}&=-N\frac{16\pi G}{3\omega}q^{-1/2}p+\frac{16N\pi{G}\pi^{Y}}{3\omega^{2}q^{1/2}}(1+\omega \check{Y}),
\end{align}
where $\mathfrak{X}_1$ is independent of $N$. Thus, $\mathcal{F}_{1}$ takes the form:
\begin{align}
\mathcal{F}_{1}&=\int_{\Sigma}d^3x(ND^{a}M-MD^{a}N)D_{a}[\frac{\omega q^{1/2}}{8\pi G}(\frac{16\pi{G}\pi^{Y}}{3\omega^{2}q^{1/2}}(1+\omega \check{Y})-\frac{16\pi G}{3\omega}q^{-1/2}p)]\nonumber\\
&=\int_{\Sigma}d^3x(ND^{a}M-MD^{a}N)D_{a}[\frac{2\pi^{Y}}{3\omega}(1+\omega \check{Y})-\frac{2}{3}p].
\end{align}
The second part is $\mathcal{F}_{2}\equiv\int_{\Sigma}d^3x(\delta\!C[N]/{\delta\!Y_{c}(x)})(\delta\!C[M]/{\delta\!\pi^{c}(x)})-(\delta\!C[N]/{\delta\!\pi^{c}(x)})(\delta\!C[M]/{\delta\!Y_{c}(x)})$. Its terms can be calculated as:
\begin{align}
	\frac{\delta\!C[N]}{\delta\!Y_{c}(x)}&=4\rho q^{1/2}D_{a}(N\Omega^{ac})+NY^c(q^{ab}Y_{a}Y_{b}-\check{Y})^{-1/2}D_{d}\pi^{d}\nonumber\\
	&=4\rho q^{1/2}\Omega^{ac}D_{a}N+N[4\rho q^{1/2}D_{a}\Omega^{ac}+Y^c(q^{ab}Y_{a}Y_{b}-\check{Y})^{-1/2}D_{d}\pi^{d}],\\
	\frac{\delta\!C[N]}{\delta\!\pi^{c}(x)}&=-\frac{N\pi_{c}}{4\rho q^{1/2}}-D_{c}[N(q^{ab}Y_{a}Y_{b}-\check{Y})^{1/2}]\nonumber\\
	&=-(q^{ab}Y_{a}Y_{b}-\check{Y})^{1/2}D_{c}N+N[-\frac{\pi_{c}}{4\rho q^{1/2}}-D_{c}(q^{ab}Y_{a}Y_{b}-\check{Y})^{1/2}].
\end{align}
Thus, $\mathcal{F}_{2}$ takes the form:
\begin{align}
	\mathcal{F}_{2}&=-\int_{\Sigma}d^3x4\rho q^{1/2}\Omega_{dc}(q^{ab}Y_{a}Y_{b}-\check{Y})^{1/2}(D^{d}ND^{c}M-D^{d}MD^{c}N)\nonumber\\
	&+\int_{\Sigma}d^3x(ND^{e}M-MD^{e}N)\Big[\pi^{c}\Omega_{ec}+4\rho q^{1/2}D^{c}[\Omega_{ec}(q^{ab}Y_{a}Y_{b}-\check{Y})^{1/2}]-Y_eD_{d}\pi^{d}\Big]\nonumber\\
	&=\int_{\Sigma}d^3x(ND^{a}M-MD^{a}N)(\pi^{b}\Omega_{ab}-Y_aD_{b}\pi^{b}).
\end{align}
The third part is $\mathcal{F}_{3}\equiv\int_{\Sigma}d^3x(\delta\!C[N]/{\delta\!q_{cd}(x)})(\delta\!C[M]/{\delta\!p^{cd}(x)})-(\delta\!C[N]/{\delta\!p^{cd}(x)})(\delta\!C[M]/{\delta\!q_{cd}(x)})$. Its terms can be calculated as:
\begin{align}
	\frac{\delta\!C[N]}{\delta\!q_{cd}(x)}&=-q^{1/2}[D^{c}D^{d}(N\frac{(1+\omega \check{Y})}{16\pi G})-q^{cd}D_{e}D^{e}(N\frac{(1+\omega \check{Y})}{16\pi G})]\nonumber\\
	&-\frac{\omega}{8\pi G}q^{1/2}D_{(a}ND_{b)}{\check{Y}}(\frac{1}{2}q^{cd}q^{ab}-q^{ac}q^{bd})+N\mathfrak{X}^{cd}\nonumber\\
	&=-q^{1/2}\frac{(1+\omega \check{Y})}{16\pi G}[D^{c}D^{d}N-q^{cd}D_{e}D^{e}N]+q^{1/2}\frac{\omega}{16\pi G}q^{cd}D_{e}ND^{e}\check{Y}+N\mathfrak{Y}^{cd},\\
	\frac{\delta\!C[N]}{\delta\!p^{cd}(x)}&=N\frac{16\pi Gq^{-1/2}}{1+\omega \check{Y}}(2p_{cd}-\frac{2pq_{cd}}{3})-N\frac{16\pi G}{3\omega}q^{-1/2}q_{cd}\pi^{Y},
\end{align}
where $\mathfrak{X}^{cd}$ and $\mathfrak{Y}^{cd}$ are independent of $N$. Thus, $\mathcal{F}_{3}$ takes the form:
\begin{align}
\mathcal{F}_{3}&=\int_{\Sigma}d^3x(MD^aD^bN-ND^aD^bM)\Big[-(2p_{ab}-\frac{2pq_{ab}}{3})-q_{ab}\frac{2}{3}\frac{(1+\omega \check{Y})}{\omega}\pi^{Y}\Big]\nonumber\\
&+(MD^aN-ND^aM)q^{1/2}\frac{\omega}{16\pi G}q^{cd}D_{a}\check{Y}\Big[\frac{16\pi Gq^{-1/2}}{1+\omega \check{Y}}(2p_{cd}-\frac{2pq_{cd}}{3})-\frac{16\pi G}{3\omega}q^{-1/2}q_{cd}\pi^{Y}\Big]\nonumber\\
&=\int_{\Sigma}d^3x(ND^aM-MD^aN)\Big[D^{b}[-(2p_{ab}-\frac{2pq_{ab}}{3})-q_{ab}\frac{2}{3}\frac{(1+\omega \check{Y})}{\omega}\pi^{Y}]+\pi^{Y}D_{a}\check{Y}\Big].
\end{align}
Collecting the above results, we obtain:
\begin{align}
\{C[N],C[M]\}&=\int_{\Sigma}d^3x(ND^{a}M-MD^{a}N)\Big[D_{a}[\frac{2\pi^{Y}}{3\omega}(1+\omega \check{Y})-\frac{2}{3}p]+\pi^{b}\Omega_{ab}-Y_aD_{b}\pi^{b}\nonumber\\
&-D^{b}[(2p_{ab}-\frac{2pq_{ab}}{3})+q_{ab}\frac{2}{3}\frac{(1+\omega \check{Y})}{\omega}\pi^{Y}]+\pi^{Y}D_{a}\check{Y}\Big]\nonumber\\
&=\int_{\Sigma}d^3x(ND^{a}M-MD^{a}N)[-2D^{b}p_{ab}+\pi^{Y}D_{a}\check{Y}+\pi^{b}\Omega_{ab}-Y_aD_{b}\pi^{b}]\nonumber\\
&=\int_{\Sigma}d^3x(ND^{a}M-MD^{a}N)C_{a}.
\end{align}

\twocolumngrid

\end{document}